\documentclass[10pt,twocolumn,twoside]{IEEEtran}

\usepackage{amsmath,amsfonts,amsthm,amssymb,color}
\usepackage{fancybox}
\usepackage[colorlinks=true]{hyperref}
\usepackage[procnumbered,ruled,vlined,linesnumbered]{algorithm2e}
\usepackage{float}
\usepackage{graphicx}
\usepackage{subcaption}
\usepackage{thmtools}
\usepackage{thm-restate}
\usepackage{array}
\usepackage{balance}

\newtheorem{theorem}{Theorem}[section]
\newtheorem{corollary}[theorem]{Corollary}
\newtheorem{lemma}[theorem]{Lemma}

\newtheorem{proposition}[theorem]{Proposition}

\newtheorem{definition}[theorem]{Definition}

\newtheorem{problem}{Problem}

\SetKw{Continue}{continue}

\newenvironment{fminipage}%
  {\begin{Sbox}\begin{minipage}}%
  {\end{minipage}\end{Sbox}\fbox{\TheSbox}}

\SetKwRepeat{Do}{do}{while}
\DontPrintSemicolon
\SetKw{KwAnd}{and}
\SetProcNameSty{textsc}
\SetFuncSty{textsc}
\SetKwInOut{Input}{Input}
\SetKwInOut{Output}{Output}

\def\Approx#1{\approx_{#1}}

\def\pr#1{\mbox{Pr}\left( #1 \right)}

\def\defeq{\stackrel{\mathrm{def}}{=}}

\def\eps{\epsilon}

\def\trace#1{\mathrm{Tr} \left(#1 \right)}

\def\Scal{\mathcal{S}}

\def\Ecal{\mathcal{E}}
\def\Fcal{\mathcal{F}}
\def\Vcal{\mathcal{V}}
\def\Gcal{\mathcal{G}}
\def\Lcal{\boldsymbol{\mathcal{L}}}
\def\Scal{\mathcal{S}}
\def\Ncal{\mathcal{N}}
\def\Mcal{\mathcal{M}}
\def\Rcal{\boldsymbol{\mathcal{R}}}
\def\Hcal{\mathcal{H}}

\newcommand{\Ima}{\text{Im}}

\newcommand\PPi{\boldsymbol{\Pi}}

\newcommand\ppi{\boldsymbol{\pi}}

\newcommand\bb{\boldsymbol{\mathit{b}}}

\newcommand\dd{\boldsymbol{\mathit{d}}}
\newcommand\ee{\boldsymbol{\mathit{e}}}

\newcommand\pp{\boldsymbol{\mathit{p}}}
\newcommand\qq{\boldsymbol{\mathit{q}}}

\newcommand\rr{\boldsymbol{\mathit{r}}}

\def\tt{\boldsymbol{\mathit{t}}}

\newcommand\vv{\boldsymbol{\mathit{v}}}

\newcommand\yy{\boldsymbol{\mathit{y}}}
\newcommand\zz{\boldsymbol{\mathit{z}}}
\newcommand\xx{\boldsymbol{\mathit{x}}}

\renewcommand\AA{\boldsymbol{\mathit{A}}}

\newcommand\DD{\boldsymbol{\mathit{D}}}
\newcommand\EE{\boldsymbol{\mathit{E}}}

\newcommand\II{\boldsymbol{\mathit{I}}}

\newcommand\KK{\boldsymbol{\mathit{K}}}

\newcommand\LL{\boldsymbol{\mathit{L}}}

\newcommand\PP{\boldsymbol{\mathit{P}}}
\newcommand\QQ{\boldsymbol{\mathit{Q}}}
\newcommand\RR{\boldsymbol{\mathit{R}}}

\newcommand\WW{\boldsymbol{\mathit{W}}}

\newcommand\XX{\boldsymbol{\mathit{X}}}

\newcommand{\zero}{\mathbf{0}}
\newcommand{\one}{\mathbf{1}}

\newcommand{\rvs}[1]{{#1}}

\DeclareMathOperator*{\argmin}{arg\,min}
\DeclareMathOperator*{\argmax}{arg\,max}

\newcommand{\kh}[1]{\left(#1\right)}

\newcommand{\svs}{P}

\newcommand{\CSL}{Absolute Leader}
\newcommand{\PSL}{Influenced Leader}

\newcommand{\csl}{absolute leader}
\newcommand{\psl}{influenced leader}

\newcommand{\ps}{influenced}

\let\oldnl\nl
\newcommand{\nonl}{\renewcommand{\nl}{\let\nl\oldnl}}

\begin{document}

\title{Shifting Opinions in a Social Network Through Leader Selection}

\author{Yuhao~Yi,~Timothy Castiglia,~Stacy~Patterson
\thanks{Y. Yi, T. Castiglia, and S. Patterson are with the Department of Computer Science, Rensselaer Polytechnic Institute, Troy, NY, 12180, USA (email: yiy3@rpi.edu, castit@rpi.edu, sep@cs.rpi.edu).}
}

\markboth{}
{Yi \MakeLowercase{\textit{et al.}}: Shifting the Opinions in a Social Network through Leader Selection}

\maketitle

\begin{abstract}
We study the French-DeGroot opinion dynamics in a social network with two polarizing parties. We consider a network in which the leaders of one party are given, and we pose the problem of selecting the leader set of the opposing party so as to shift the average opinion to a desired value. When each party has only one leader, we express the average opinion in terms of the transition matrix and the stationary distribution of random walks in the network. The analysis shows balance of influence between the two leader nodes. We show that the problem of selecting at most $k$ absolute leaders to shift the average opinion is  $\mathbf{NP}$-hard. Then, we reduce the problem to a problem of submodular maximization with a submodular knapsack constraint and an additional cardinality constraint and propose a greedy algorithm with upper bound search to approximate the optimum solution. We also conduct experiments in random networks and real-world networks to show the effectiveness of the algorithm.
\end{abstract}%


%
\IEEEpeerreviewmaketitle

\section{Introduction}

Social networks have become increasingly influential in shaping public opinions. 
Within this field,  the problem of  designing mechanisms to effectively shift opinions in a social network has received great interest in last two decades~\cite{YOASS13,GTT13,VFFO14,HZ18,AKPT18,MMT18,MA19,MP19}. Much of the existing work studies the problem of choosing individuals in the network to be opinion leaders so as to maximize the influence of a particular opinion, for example, to shift the average opinion of the network to an extreme opinion. However, fine-grained optimization of the average opinion has not been well studied.

In this paper, we study the problem of shifting the average opinion of a network to a given value, which generalizes the intensely studied influence maximization problem. We consider a continuous-time French-DeGroot opinion model with two polarizing parties. 
The French-DeGroot model~\cite{Fre56,DeG74} is one of the most popular models for opinion dynamics.
In the model, the social network is represented by a graph, with nodes corresponding to individuals.
Each node has a  real scalar-valued state that represents the individual's opinion. 
Each node updates its state continuously by comparing its state and the states of its neighbors. 
We consider a variation on this model where the nodes consists of leaders nodes, defined as the nodes with external reference values, and follower nodes, defined as those without external information.

\rvs{
We assume that there are two opposing sources of opinion, $1$ and $0$. These could represent, for example, support for Party A (1) or Party B (0), or these sources could represent support and opposition to an event or outcome. 
In the model we adopt, all members of the social network take opinion values in the interval $[0,1]$. 
A firm supporter of party A (or pro-event individual) has an opinion close to  $1$, and a unquestioned supporter of Party B (or anti-event individual) holds an opinion close to $0$. 
Individuals with opinion $\frac{1}{2}$ are considered as completely neutral.}

Each party controls a set of nodes as their opinion leaders. 
The leader nodes can be fully or partially controlled by each party.  If a leader node is fully controlled, its opinion is set to the constant opinion value of that party and never changes over time. \rvs{We call these leaders \emph{\csl s} and call such a system an \rvs{\emph{absolute leader system}}.
If a leader node is partially controlled, it receives a constant input from the corresponding party as a reference value,  and it adjusts its state according to the reference value and the states of its neighbors.  One can think of these leaders as being influenced through a relationship with an individual that is a direct source of the opinion but is not part of the network, i.e., an external party leader.
We define such partially controlled nodes as \emph{influenced leaders}, and we call this kind of system an \emph{influenced leader systems}.}

\rvs{We consider two leader selection problems, one for each type of system. In both cases, we assume that the network already has a leader set for party 0, and our goal is to identify the leader set for party 1 so as to shift the average opinion of the social network towards a target value. 
In the \emph{absolute leader system}, this translates to selecting individuals in the network to act as \emph{\csl s}, for example, by hiring them into the party. In the \emph{influenced leader system}, this leader selection translates to forming relationships between the identified set of \psl s within the network and the external party leader with opinion 1.}

\rvs{Our problem formulation is related to the well studied problem of influence maximization~\cite{GTT13,VFFO14,HZ18, MA19}, i.e.,  maximizing the average opinion of the network by choosing leaders for party 1, while the leaders for party 0 are fixed. However, there are cases where maximizing the average opinion is not beneficial. It is well known that a large group of people tends to have polarizing opinions, and the problem of depolarizing the opinions in a group of interacting individuals has received interest in social psychology~\cite{Sun99}. In this case, one may seek to balance the network opinion around a target value of $\frac{1}{2}$. Moreover, the opinion of an individual relates to his or her behaviors~\cite{Fri10, Fri15}. In particular, it can be viewed as the probability that a user adopts a behavior. From this perspective, party 1 can
achieve a desired level of participation in a voluntary activity in a large network by shifting the average opinion to a certain target value.}

We begin by analyzing the two proposed models, and we propose the concept of domination score to characterize the balance of influence between leaders of two parties. This analysis relates the models to \rvs{properties of random walks in a network}.  We also identify the optimal solution to the leader selection problem for each model when a single leader is chosen for each party.
Next, we study the general problem of choosing a leader set for party 1 with a given cardinality, when the leader set for party 0 is already identified. For \rvs{absolute leader systems}, we prove the $\mathbf{NP}$-hardness of the problem by a reduction from the vertex cover problem on $3$-regular graphs. We also show the monotonicity and submodularity of the average steady-state opinion as a function of the leader set of party $1$, for both \rvs{absolute and influenced} leader systems. 
Then, we propose an algorithm for the leader selection problems with provable  approximation guarantees. Our algorithm finds an appropriate upper bound for a greedy routine that \rvs{approximately} solves a submodular cost submodular knapsack (SCSK) problem \rvs{with an additional cardinality constraint. we are not aware of any previous work on SCSK problems with cardinality constraints.}

\subsubsection*{Related work}

In the last two decades, many works considered the  French-DeGroot model  with leaders accessing the same reference value~\cite{BH06,PB10,LFJ14,CBP14,CABP14}. In such systems, leader selection problems have been formulated for different objectives such as minimizing the convergence error~\cite{CABP14} or minimizing the total deviation from the reference value of the system in  the  presence of additional noise on followers~\cite{PB10,LFJ14,CBP14}. These combinatorial optimization problems
are often intractable. For example, the leader selection problem proposed in~\cite{PB10} has been proven to be {\bf{NP}}-hard~\cite{LPS+18}. Various approaches have been proposed to address these problems, including convex relaxation heuristics~\cite{LFJ14} and greedy algorithms~\cite{CABP14,MP18} with constant approximation ratios. 

Another line of works consider leaders with different reference values, in particular 
two group of leaders with polarizing opinions. 
In this case, the steady-state opinions of all nodes fall into the interval 
of leader states~\cite{ACFO13,CF16}. 
In such systems, different leader selection problems have also been studied.~\cite{VFFO14} investigated the problem of single leader placement to maximize its influence. 
The work \cite{MA19} studied a problem of choosing leaders to maximize influence of the leader set in a French-DeGroot model where leaders have specified stubbornness, and~\cite{HZ18} investigated a similar maximization problem. Both works proved the monotonicity  and submodularity of the average opinion in a French-DeGroot opinion network with \rvs{influenced} leader dynamics. 
\cite{GTT13} studied the influence maximization problem in the Friedkin-Johnsen model, which is related to a French-DeGroot opinion network with \rvs{absolute} leaders in special cases but not equivalent, in general. This work proved the submodularity of the average opinion in their model as a function of leader nodes and the {\bf{NP}}-hardness of the average opinion maximization problem. Typical greedy algorithms were applied to these problems due to submodularity of the objective functions. 
In contrast, our work studies the problem of shifting the average opinion of the network to any specified value. Our problem thus includes the influence maximization problem as a special case. In addition, we show that our problem cannot be directly treated as submodular maximization problem with a cardinality constraint. Thus, a more sophisticated optimization algorithm is needed.

\subsubsection*{Paper outline}
The reminder of the paper is organized as follows. In Section~\ref{prepre.sec}, we introduce basic notations and concepts. In Section~\ref{pre.sec}, we present the system model and the problem formulation. In Section \ref{analysis.sec}, we give an explicit form of the steady-state opinion 
vector using the Laplacian of an augmented graph, and we show how this relates to the balance of the leader nodes' influence  in a network. We also prove the hardness of the investigated problem in \rvs{influenced} leader systems. In Section~\ref{algorithm.sec}, we propose a greedy algorithm with an upper bound search and provide provable bounds on the approximation ratio of the algorithm.  Section~\ref{experiments.sec}  presents experimental results. Finally, we conclude in Section~\ref{conclusion.sec}.

\section{Preliminaries}
\label{prepre.sec}
In this section, we introduce the notation of a graph and its matrix representations. Further, we review the concepts of \rvs{hitting time, commute time,} resistance distance, and information centrality, which are used as analytical tools in this paper.

\paragraph*{Vectors and Matrices}
We use $\ee_u$ to denote the $u$-th canonical basis  vector of $\mathbb{R}^n$. The vector $\bb_{u,v}$ is defined as $\bb_{u,v}\defeq \ee_u-\ee_v$. $\one_n$ represents the all-one vector with length $n$, and $\zero_n$ ($\zero_{p\times q}$) represents the all-zero vector (or matrix) with legnth $n$ (or size $p\times q$). We also use these notations without specifying the sizes if they are implied in context. Apart from these exceptions ($\ee_u$, $\bb_{u,v}$, $\one_n$ and $\zero_n$), a vector or matrix with subscripts denotes the vector or  submatrix with indices specified by the subscripts. 
 For example, given a vector $\xx$, $\xx_i$ is its $i$-th entry, and $\xx_{\mathcal{I}}$ is a vector consisting of entries $\xx_i$ for all $i\in\mathcal{I}$. For a matrix $\XX$, $\XX_{i,j}$ is the $(i,j)$-{th} entry of $\XX$ and $\XX_{\mathcal{I},\mathcal{J}}$ is the submatrix of $\XX$ consisting of the entries of $\XX$ whose rows are in ${\mathcal{I}}$ and columns are in ${\mathcal{J}}$. In addition, we use $\II$ to denote the identity matrix,  and we use $\XX^\dag$ to denote the
Moore Penrose pseudoinverse  of the matrix $\XX$. 

\paragraph*{Graphs and their Matrix Representations}
\rvs{We denote a directed graph as $\Gcal = (\Vcal, \Ecal, w)$, where $\Vcal$ and $\Ecal$ are the node set and edge set of the graph, respectively, with $|\Vcal| = n$ and $|\Ecal|= m$. An undirected graph can be viewed as a symmetrically coupled bidirectional graph in this context. 
We let $e=(u,v)\in\Ecal$ represent an edge from nodes $u$ to node $v$, and $w: \Ecal\to \mathbb{R}^+$ is the edge weight function.   We denote $\Ncal_v^\downarrow$ as the set of in-neighbors of $v$ ($u\in \Ncal_v^\downarrow$ iff $(u,v)\in\Ecal$), and $\Ncal_v^\uparrow$ as the set of out-neighbors of $v$ ($u\in \Ncal_v^\uparrow$ iff $(v,u)\in\Ecal$).} In addition, for a graph $\Gcal = (\Vcal, \Ecal, w)$, and a subset of nodes $V\subseteq \Vcal$, 
we denote the subgraph supported on $V$ as $\Gcal[V]=(V,E,\omega)$, where $E = \{e=(u,v)\in \Ecal: u,v\in V\}$ and $\omega(e)=w(e)$ for all $e\in E$. Further, we define the plus operation on graphs as follows. For two graphs $\Gcal_1=(\Vcal_1, \Ecal_1, w_1)$ and $\Gcal_2=(\Vcal_2, \Ecal_2, w_2)$, let $\Hcal=(\mathcal{U},\Mcal, \omega) = \Gcal_1 + \Gcal_2$  be a new graph with  $\mathcal{U} = \Vcal_1\cup \Vcal_2$, $\Mcal = \Ecal_1\cup \Ecal_2$, and $\omega:\Mcal \to \mathbb{R}^+$  the new edge weight function defined as $\omega(e)= w_1(e)$ if $e\in (\Ecal_1\backslash \Ecal_2)$, $\omega(e)= w_2(e)$ if $e\in (\Ecal_2\backslash \Ecal_1)$, and $\omega(e)= w_1(e)+w_2(e)$ if $e\in (\Ecal_2\cap \Ecal_1)$. 

The weighted Laplacian matrix of a graph is defined as \rvs{$\LL\defeq \DD -\AA$}, where $\AA$ is the adjacency matrix with ${\AA_{u,v} = w(e)}$ for $e=(u,v)\in \Ecal$ and $\AA_{u,v} = 0$ for $(u,v)\notin \Ecal$, and $\DD$ is the \rvs{out-degree} diagonal matrix, where $\DD_{u,u} = \sum_v \AA_{u,v}$ and $\DD_{u,v}=0$ if $u\neq v$. 
From the definition, it is clear that \rvs{$\LL = \sum_{(u,v)\in\Ecal}w(u,v)\bb_{u,v}\ee_{u}^{\top}$}.

 \rvs{For a matrix (vector, scalar) associated with a graph, we sometimes use a superscript to explicitly show that it corresponds to the graph. For example, $\LL^{\Gcal}$ is the Laplacian matrix of graph $\Gcal$.}
\rvs{
\paragraph*{Random walks on graphs}
We define $\WW \defeq \AA^\top \DD^{-1}$ as the random walk \emph{transition matrix} of graph $\Gcal$. A random walker has a probability $\WW_{i,j} = \frac{\AA_{j,i}}{\DD_{j,j}}$ to transition from vertex $j$ to vertex $i$.
When the graph is strongly connected there exists a positive vector (unique up to scaling) such that $\ppi = \WW \ppi$. When the vector is scaled such that $\sum_{v} \ppi_v =1$, $\ppi$ is called the \emph{stationary distribution} of the random walk defined by $\WW$. We define $\PPi$ as a diagonal matrix in which $\PPi_{v,v} \defeq \ppi_v$ for all vertex $v\in\Vcal$. We note that $\LL = \DD(\II - \WW^{\top})$ and $\LL\one = \zero$. 

In a connected graph $\Gcal$, the \emph{hitting time} from vertex $u$ to $v$ is the expected number of steps that a random walker, starting from vertex $v$, takes until it hits $u$ for the first time. We denote by $H_{u,v}$ the hitting time from $u$ to $v$.
\begin{lemma}[Hitting time~\cite{CKPP+16a,CKPP+16b}] 
\label{hitting.def}
In a strongly connected graph $\Gcal$,
\begin{align*}
H^\Gcal_{u,v} = \bb_{u,v}^\top (\II - (\WW)^\top)^{\dag} \PPi^{-1} (\ppi - \ee_{v})\,.
\end{align*}
If $\Gcal$ is an undirected graph, $H^\Gcal_{u,v} = 2m\cdot \bb_{u,v}^\top \LL^{\dag} (\ppi - \ee_{v})$.
\end{lemma}
The \emph{commute time} $C_{u,v}$ is defined as $C_{u,v}\defeq H_{u,v} + H_{v,u}$.
\begin{lemma}[Commute time~\cite{CKPP+16a,CKPP+16b}] 
\label{commute.def}
In a strongly connected graph $\Gcal$,
\begin{align*}
C^{\Gcal}_{u,v} = \bb_{u,v}^\top (\II - (\WW)^\top)^{\dag} \PPi^{-1}\bb_{u,v}\,.
\end{align*}
If $\Gcal$ is an undirected graph, $C^\Gcal_{u,v} = 2m\cdot \bb_{u,v}^\top \LL^{\dag} \bb_{u,v}$.
\end{lemma}

}

\paragraph*{Effective Resistance and Information Centrality}
Given an undirected graph $\Gcal$, we define an \emph{electrical network} $\overline{\Gcal}$. In $\overline{\Gcal}$, every edge $e$ of $\Gcal$ is replaced by a resistor of resistance $1/w(e)$, and the resistors are connected if the edges are incident. Then, the \emph{effective resistance} between node $u$ and $v$ in graph $\Gcal$ (or electrical graph $\overline{\Gcal}$) is defined as the voltage difference between vertices $u$ and $v$ in $\overline{\Gcal}$ when unit current is injected from $u$ and extracted from $v$. We recall the following lemma relating to effective resistance.
\begin{lemma}[Effective Resistance~\cite{KR93}]
\label{resistance.def}
In a connected undirected electrical network defined by $\Gcal = (\Vcal, \Ecal, w)$, the effective resistance between nodes $u$ and $v$ is 
\begin{align*}
R^{\Gcal}_{u,v} =(\LL^\dag)_{v,v} - 2(\LL^\dag)_{v,u}+(\LL^\dag)_{u,u}\,.
\end{align*}
\end{lemma}
We further recall the related definition of information centrality.
\begin{definition}[Information Centrality~\cite{SZ89}]
In a connected undirected graph $\Gcal = (\Vcal, \Ecal, w)$, the information centrality of a vertex $u$ is defined by 
\begin{align*}
\theta^{\Gcal}(u) = \frac{n}{\sum_{v\in\Vcal}R^{\Gcal}_{u,v}}\,.
\end{align*}
\end{definition}
From  Lemma~\ref{resistance.def} we obtain
\begin{align*}
\sum_{u\in\Vcal} R^{\Gcal}_{u,v} = n\cdot \kh{\LL^{\dag}}_{v,v} + \trace{\LL^{\dag}}\,.
\end{align*}

\section{Problem Formulation}
\label{pre.sec}

We consider a \rvs{directed strongly connected} graph ${\Gcal = (\Vcal, \Ecal, w)}$.
Nodes represent individuals in the social network, \rvs{and an edge $(u,v)\in\Ecal$ models a social link from node $u$ to node $v$, indicating that node $u$ follows node $v$, or node $v$ exerts influence on node $u$.} Edge weights represent the strengths of the social links. 
Each node $v$ has a scalar-valued state $\xx_v \in \mathbb{R}$, which represents its opinion. The node set can be divided into a leader set $S$ and a follower set $F$. The set $S$ can be further divided into two disjoint sets $S_0$ and $S_1$, which are sets of nodes controlled by two parties, namely party $0$ and party $1$. All \rvs{nodes in $S_0$ have} access to reference value $0$, and all \rvs{nodes in $S_1$} have access to reference value $1$. \rvs{Nodes in $F$} update their states according to a diffusion law.

\subsection{System Dynamics}
We consider the French-DeGroot opinion model with \rvs{\csl s and a variation of this model with \psl s that are connected to external absolute sources of information}. The two models differ in how the leaders use their reference values.

In the \rvs{absolute} leader system,
leaders initialize their states with $0$ (for $v\in S_0$) or $1$ (for $v\in S_1$), and their states remain unchanged over time.
The dynamics of a leader node $v$ is characterized by $\dot{\xx}_v(t)=0$. A follower node $v$ begins with an arbitrary initial state $\xx_v(0) = \xx_v^0$, and it updates its state by the dynamics
\begin{align*}
\dot{\xx}_v(t) = -\sum_{u\in \Ncal^\uparrow_v}w(v,u)\kh{\xx_v(t) - \xx_u(t)}\,.
\end{align*}

We  partition the state vector $\xx$ as
$$\xx = \kh{\xx_S^\top \,\,\xx_F^\top}^\top\,$$
where $\xx_S$ is associated with the leaders and $\xx_F$ is associated with the followers. Similarly, we partition the Laplacian matrix $\LL^{\Gcal}$ and adjacency matrix $\AA$ into blocks as 
\begin{align*}
\LL^{\Gcal} = \begin{pmatrix}
\LL_{S,S} & \LL_{S,F}\\
\LL_{F,S} & \LL_{F,F}
\end{pmatrix}\,.
\end{align*}
Then, the dynamics of the leaders and the followers can be written as
\begin{align}
\label{sys1Leaders.sys}
\dot{\xx}_S(t) & = \zero\\
\label{sys1Followers.sys}
\dot{\xx}_F(t) & = -\LL_{F,F}\xx_F - \LL_{F,S}\xx_S\,.
\end{align}

In the system described by (\ref{sys1Leaders.sys}) and (\ref{sys1Followers.sys}), the steady-state values of the leader nodes are
\begin{align}
\hat{\xx}_S = \xx_S^0
\end{align}
for $v \in S$. 
Since $-\LL_{F,F}$ is Hurwitz  for a non-empty leader set $S$~\cite{CRE12}, the system converges to a single stable steady-state~\cite{ME10}. Letting $\dot{\xx}_F(t) = \zero$,  we obtain the steady-state of the followers 
\begin{align}
\label{steady.eqn}
\hat{\xx}_F = -(\LL_{F,F})^{-1}\LL_{F,S}\hat{\xx}_{S}\,.
\end{align}
We note that $\LL_{F,S}\hat{\xx}_S$ 
can be viewed as the sum of columns of $\LL_{F,S}$ that correspond to $1$-leaders (columns of $0$-leaders are weighted by $0$).

\rvs{In the influenced leader system, disjoint subsets of nodes $S_0$ and $S_1$ are influenced by two external party leaders with opinions 0 and 1, respectively. 
These external nodes are not part of the graph $\Gcal$, and further, they do not change their opinions. 
Each of the \psl s in $S_0 \cup S_1$ updates its state according to its current state, the states of its neighbors, and the reference value from its external leader, 0 for nodes in $S_0$ and 1 for nodes in $S_1$.} 

The system can start from any initial state and the dynamics is given by
\begin{align*}
\dot{\xx}_v\! &=\! -\!\!\sum_{u\in \Ncal^{\uparrow}_v}\!\!w(v,u)\!\kh{\xx_v(t) - \xx_u(t)} + \kappa_v\kh{0-\xx_v(t)}\!,  v\in S_0,\\
\dot{\xx}_v\! &=\! -\!\!\sum_{u\in \Ncal^{\uparrow}_v}\!\!w(v,u)\!\kh{\xx_v(t) - \xx_u(t)} + \kappa_v\kh{1-\xx_v(t)}\!, v\in S_1,\\
\dot{\xx}_v\! &=\! -\!\!\sum_{u\in \Ncal^{\uparrow}_v}\!\!w(v,u)\!\kh{\xx_v(t) - \xx_u(t)}\!, \quad v\in F\,.
\end{align*}
where the value $\kappa_v$ is the weight that \rvs{the \psl} puts on its reference value. We also refer to it as the \emph{stubbornness} of the node. The dynamics can be expressed more compactly as
\begin{align}
\label{sys2.sys}
\dot{\xx} = -\kh{\LL^{\Gcal} + \EE^S \KK} \xx + \EE^{S_1}\KK\one\,,
\end{align}
where $\EE^S$ is the diagonal matrix with $\EE^S_{v,v}=1$ for $v\in S$ and  $\EE^S_{v,u}=0$ otherwise; $\EE^{S_1}$ is defined similarly with non-zero entries for $v\in S_1$. The matrix $\KK$ is diagonal with $\KK_{v,v} = \kappa_v$, the stubbornness of vertex $v$ if chosen as an \rvs{\ps} node. 

For system (\ref{sys2.sys}), $-(\LL^{\Gcal} + \EE^S \KK)$ is Hurwitz for a non-empty leader set $S$, so the system converges to a single steady-state.  We let $\dot{\xx}(t)=0$ and obtain the steady-state values of all nodes
\begin{align}
\label{partialSteady.eqn}
\hat{\xx} = \kh{\LL^{\Gcal} + \EE^S \KK}^{-1}\EE^{S_1}\KK\one\,.
\end{align}
In this paper, we study the average opinion of all nodes in the network. 
\begin{definition}
\label{muFunction.def}
In  \rvs{both the absolute and influenced leader systems}, given the leader set $S_0$, the \emph{average opinion} $\mu$ of a network  as a function of leader set $S_1$ is defined as
\begin{align}
\mu(S_1) \defeq \frac{1}{n} \sum_{v\in\Vcal} \hat{\xx}_v\,.
\end{align}
\end{definition}

Besides the above definition, $\mu(S_1)$ has an interesting interpretation in an opinion-behavior model based on the French-DeGroot model. We can model the opinion-behavior linkage in the system by 
treating $\hat{\xx}_v$ as the success probability of a Bernoulli random variable $X_v$ of taking the value $1$. 
In the social network,  $X_v=1$ indicates the event that node (individual) $v$ \rvs{takes an action}, and $X_v=0$ indicates the event that $v$ \rvs{does not take an action.}
We recall that $n = |\Vcal|$ for both the \rvs{absolute} and \rvs{influenced} leader systems.  We then assume $X_1, X_2, \ldots, X_v, \ldots, X_n$ to be $n$ mutually independent Bernoulli random variables associated with corresponding nodes in the network. In particular $X_v$ is defined by
\begin{align}
\pr{X_v=1} &= \hat{\xx}_v \,,\nonumber\\
\pr{X_v=0} &= 1 - \hat{\xx}_v\,, \nonumber
\end{align}
for all $v\in \Vcal$, and therefore  $E[X_v] = \hat{\xx}_v$.

We are interested in the fraction of nodes that \rvs{take an action}. We define the random variable $\overline{X}:=\frac{1}{n}\sum_v {X_v}$. Since $X_v$ are independent bounded random variables, $\overline{X}$ concentrates at
\begin{align}
\mu(S_1)=\frac{1}{n}\sum_v \hat{\xx}_v. 
\end{align}


According to the Hoeffding's inequality,
\begin{align}
\pr{|\overline{X}-\mu(S_1)| \geq \sqrt{\frac{\ln n}{n}}\,} \leq \frac{2}{n^2}\,,
\end{align}
which indicates that $\mu(S_1)$ determines \rvs{ the fraction of the population that take an action} in a large network, with a diminishing error bound and a diminishing probability that this bound is violated. \rvs{Therefore, a party can control the fraction of population that take part in an activity or event  by shifting the average opinion of the network to a certain value. 
}

\subsection{Leader Selection Problems}
\rvs{In a system where  the set $S_0$ is given, we define the problem of choosing at most $k$ leaders for the set $S_1$,}
such that the average opinion of all nodes (including leaders and followers) $\mu(S_1)$  is closest to a given value $\alpha$. 
\rvs{Specifically, we are interested in minimizing the following objective function,
\begin{align}
\label{objective.eqn}
f(P, \alpha)\defeq |\mu(P) - \alpha|\,.
\end{align}
}
\rvs{We first formally define the problem for the absolute leader system.}
 \begin{problem}[\rvs{\underline{A}bsolute} \underline{L}eader \underline{S}election]
\label{stubborn.prob} 
\rvs{In an absolute leader system,} given a \rvs{strongly connected directed} graph $\Gcal=(\Vcal, \Ecal, w)$, an opinion $0$ absolute leader set $S_0\neq \emptyset$, a specified value $\alpha \in [0,1]$, a candidate set $Q\subseteq \Vcal\backslash S_0$, $|Q|=q$, and an integer $1\leq k\leq q$, find the node set $S_1\subseteq Q$, $|S_1|\leq k$ such that
\begin{align}
S_1\in \argmin_{P \subseteq Q, |P|\leq k}\rvs{ f(P, \alpha)\,}.
\end{align}
\end{problem}

We define a similar problem for the \rvs{influenced} leader system.
\begin{problem}[\rvs{\underline{I}nfluenced} \underline{L}eader \underline{S}election]
\label{partial.prob}
 \rvs{In an influenced leader system}, given a \rvs{strongly connected directed} graph ${\Gcal = (\Vcal, \Ecal, w)}$, an opinion $0$ leader set $S_0\neq\emptyset$, a stubbornness function of $0$ leader nodes $\kappa_0: S_0 \to \mathbb{R}^+$, a specified value $\alpha\in [0,1]$, a candidate set $Q\subseteq \Vcal\backslash S_0$, $|Q|=q$, another stubbornness function $\kappa_1: Q\to \mathbb{R}^+$, and a integer $1\leq k \leq q$,  find the node set $S_1\subseteq Q$, $|S_1|\leq k$ such that
\begin{align}
S_1 \in \argmin_{P \subseteq Q, |P|\leq k} \rvs{f(P, \alpha)\,}.
\end{align}
\end{problem}

We note that for both Problems~\ref{stubborn.prob} and \ref{partial.prob}, influence maximization corresponds to the degenerate case of $\alpha=1$.

\section{Analysis}
\label{analysis.sec}

In this section, we give analytical solutions for Problems~\ref{stubborn.prob} and \ref{partial.prob} for the case where $k=1$. We also present hardness results for the case where $k>1$. 

Our analysis utilizes  \rvs{a leader-equivalent} graph  to give analytical expressions for the average opinion of the network. \rvs{Furthermore, for a network with a single leader for each party, we express the average opinion using the transition matrix and the stationary distribution of random walks in the network.} 


\subsection{Opinions in \rvs{Leader-Equivalent Systems}}

We note that the dynamics of both the \rvs{absolute} leader system and the \rvs{influenced} leader system can be fully characterized by a system defined in \rvs{\emph{a leader-equivalent graph}}. For these two different kinds of systems, we construct the corresponding \rvs{leader-equivalent} graphs in different ways.
%

\begin{figure}[htbp]
\centering
\includegraphics[scale=.8]{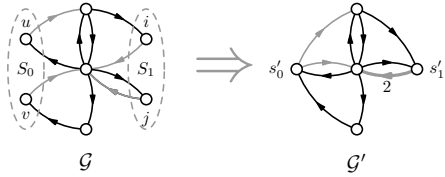}
\caption{An example of constructing \rvs{a leader-equivalent} graph \rvs{for an absolute} leader system. Nodes $u$ and $v$ in $\Gcal$ become the leader $s'_0$ in $\Gcal'$, and nodes $i$ and $j$ in $\Gcal$ become the leader $s_1'$ in $\Gcal'$. Edges without labels are weighted $1$; otherwise, edges are labeled with their weights.}
\label{aug_stub.fig}
\end{figure}
The system described by (\ref{sys1Leaders.sys}) and (\ref{sys1Followers.sys}) is equivalent to a system  in which all nodes in $S_0$ are identified as a single \rvs{\csl} $s'_0$, and all nodes in $S_1$ are identified as a single \rvs{\csl} node $s'_1$.
We denote the \rvs{contracted} graph by $\Gcal'=(\Vcal',\Ecal', w')$, where $\Vcal' = F \cup\{s'_0\}\cup\{s'_1\}$, $\Ecal' =\{(u,v): u,v\in F\}\cup \{(u,s'_0): (\Ncal_u \cap S_0)\neq \emptyset \} \cup \{(u,s'_1): (\Ncal_u \cap  S_1) \neq \emptyset\}$, and $w'(u,v)=w(u,v)$ if $u,v\in F$, $w'(u,s'_0) = \sum_{v\in (S_0\cap \Ncal_u)}w(u, v)$, and $w'(u,s'_1) = \sum_{v\in (S_1\cap \Ncal_u)}w(u, v)$. In addition, we define $S' = \{s'_0,s'_1\}$ and $F' = \Vcal' \backslash S'$. Note that $F' = F$ in this case.
Figure~\ref{aug_stub.fig} shows an example of constructing  \rvs{a leader-equivalent} graph for an \rvs{absolute} leader system.

We denote the Laplacian matrix of $\Gcal'$ as $\LL^{\Gcal'}$. Then the dynamics of $F'$ in the system defined on the \rvs{leader-equivalent} graph is expressed by
\begin{align}
\label{augComp.eqn}
\dot{\xx}_{F'}(t) & = -\LL^{\Gcal'}_{F',F'}\xx_{F'}  - \LL^{\Gcal'}_{F',\{s'_1\}}\,.
\end{align} 



\begin{figure}[htbp]
\centering
\includegraphics[scale=.8]{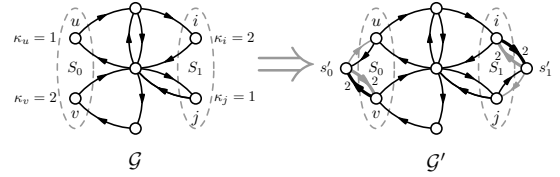}
\caption{An example of constructing a \rvs{leader-equivalent} graph from an \rvs{influenced} leader system.}
\label{aug_part.fig}
\end{figure}
The \rvs{influenced leader} system described by (\ref{sys2.sys}) is equivalent to a system  in which two virtual \rvs{absolute}  leaders $s'_0$ and $s'_1$ are added to the graph, and all nodes in the original network $\Gcal$ are treated as followers. 
We define the augmented graph as $\Gcal' = (\Vcal', \Ecal', w')$, where $\Vcal' = \Vcal \cup\{s'_0\}\cup\{s'_1\}$, and \rvs{$\Ecal' = \Ecal \cup \{(u,s'_0):u\in S_0 \} \cup \{(u,s'_1):u\in S_1\}\cup\{(s'_0,u):u\in S_0 \} \cup \{(s'_1,u):u\in S_1\}$}, $w'(u,v)=w(u,v)$ if $(u,v)\in \Ecal$, \rvs{$w'(u,s'_0) = w'(s'_0,u) =\kappa_u$} if $u\in S_0$, and \rvs{$w'(u, s'_1)=w'( s'_1,u)= \kappa_u$} if $u\in S_1$. We again define $S' = \{s'_0, s'_1\}$ and $F' = \Vcal' \backslash S'$; in this case $F'=\Vcal$.
Figure~\ref{aug_part.fig} shows an example of constructing \rvs{a leader-equivalent graph} for an \rvs{influenced} leader system. 
With this augmented graph, the dynamics of the \rvs{influenced} leader system is also described by~(\ref{augComp.eqn}).



By constructing the corresponding \rvs{leader-equivalent} graphs, we can study both \rvs{absolute} and \rvs{influenced} leader systems using a unified framework. We remark that this does not mean the systems are equivalent. Choosing leaders in different system models leads to different \rvs{leader-equivalent} graphs and hence different steady-states, although system  (\ref{sys2.sys}) approaches system (\ref{sys1Followers.sys}) as $\KK_{v,v} \to +\infty$ for all $v\in (S_0\cup S_1)$.

For both the \rvs{absolute} and \rvs{influenced}  leader systems,  the \rvs{nodes} $s'_1$ and $s'_0$ are the only nodes that directly use reference values as their states in the \rvs{leader-equivalent} graph. 
Their steady states are $\xx_{s'_0} = 0$ and $\xx_{s'_1} = 1$.
The steady states of all remaining nodes satisfy
\begin{align}
\label{remainSteady.eqn}
\LL^{\Gcal'}_{F',F'} \hat{\xx}_{F'} + \LL^{\Gcal'}_{F',\{s'_1\}}=\zero\,.
\end{align}
\rvs{We note that the edges from $s'_0$ or $s'_1$ to other nodes are not used according to the dynamics. We deliberately add these edges to make the graph strongly connected, which facilitates our analysis.}

\rvs{Let $ (\AA^{\Gcal'})^{\top}(\DD^{\Gcal'})^{-1}$ 
be the random walk matrix of a \rvs{leader-equivalent} graph $\Gcal'$. Then, we define the following matrices for $\Gcal'$:
\begin{align}
\Lcal^{\Gcal'} &\defeq \PPi (\II - (\WW^{\Gcal'})^\top)\,,\\
\Rcal^{\Gcal'} &\defeq (\II - (\WW^{\Gcal'})^\top)^{\dag} \PPi^{-1}\,.
\end{align}
In general $(\Lcal^{\Gcal'})^\dag \neq \Rcal^{\Gcal'}$, but for any $\pp\perp\one, \qq\perp\one$, $\pp^\top(\Lcal^{\Gcal'})^\dag \qq = \pp^\top \Rcal^{\Gcal'} \qq$. 
For more details we refer the readers to the full version~\cite{CKPP+16b} of \cite{CKPP+16a}. For an undirected graph, $\Lcal^{\Gcal'} = \frac{1}{2m}\LL^{\Gcal'}$. 

}

\begin{proposition}
\label{explicitProb.thm}
For either an \rvs{absolute}  leader system or an \rvs{influenced} leader system, we consider its \rvs{leader-equivalent} graph $\Gcal'$. 
For any node $v\in \Vcal'$, the steady state value $\hat{\xx}_{v}$ is given by 
\rvs{
\begin{align}
\label{explicit.eqn}
\hat{\xx}_{v} = \frac{\bb_{v,s'_0}^\top\Rcal^{\Gcal'}\bb_{s'_1,s'_0}}{\bb^\top_{s'_1,s'_0}\Rcal^{\Gcal'} \bb_{s'_1,s'_0}}= \frac{\bb_{v,s'_0}^\top(\Lcal^{\Gcal'})^{\dag}\bb_{s'_1,s'_0}}{\bb^\top_{s'_1,s'_0}(\Lcal^{\Gcal'})^{\dag}  \bb_{s'_1,s'_0}} \,.
\end{align}
When $\Gcal$ is an undirected graph, the expression degenerates to
\begin{align}
\label{degexplicit.eqn}
\hat{\xx}_{v} = \frac{\bb_{v,s'_0}^\top(\LL^{\Gcal'})^{\dag} \bb_{s'_1,s'_0}}{\bb^\top_{s'_1,s'_0}(\LL^{\Gcal'})^{\dag}  \bb_{s'_1,s'_0}} \,.
\end{align}
}
\end{proposition}
\rvs{The correctness of the result in Proposition \ref{explicitProb.thm} can be verified by plugging (\ref{explicit.eqn}) into (\ref{remainSteady.eqn}), and the uniqueness is guaranteed by the fact that $\LL^{\Gcal'}_{F',F'}$ is full rank and $\EE^{S_1} \KK' \one$ is non-zero. We leave the details to Appendix \ref{appendix2.sec}. 

The value of $\hat{\xx}_{v}$ is, in fact, the \emph{escape probability} of node $v$, which is defined as the probability that a random walker starting from vertex $v$, reaches node $s'_1$ before it reaches node $s'_0$\footnote{The references \cite{CKPP+16a,CKPP+16b} discussed the escape probability of a node in a directed graph, although these papers did not include a correct expression.},
We note that the expression (\ref{degexplicit.eqn}) was given in~\cite{CF16} in a different context. 
\cite{CF16} studied an opinion dynamics model where the sum of differences between the states of a node and its neighbors is divided by the out-degree of the node before it is applied as a negative feedback to the state of the node. If the leaders take the same values, the system studied in~\cite{CF16}  has a different convergence rate than the absolute leader system  but shares the same steady-state values. 
}

\subsection{Single Leader for Each Party}
For \rvs{absolute} leader systems, if $|S_1|=|S_0|=1$, the \rvs{leader-equivalent} graph $\Gcal'$ is the same as the original graph $\Gcal$. We let the leaders in $\Gcal$ be denoted  $s_0$ and $s_1$ for parties with opinion $0$ and $1$, respectively.  Then,  by Proposition~\ref{explicitProb.thm},
\rvs{
\begin{align}
\label{muSingleStubborn.eqn}
\mu(S_1) =\! \frac{(\Lcal^{\dag})_{s_0,s_0} - (\Lcal^{\dag})_{s_0,s_1}}
{(\Lcal^{\dag})_{s_0,s_0}\! -\! (\Lcal^{\dag})_{s_0,s_1}\!  +\! (\Lcal^{\dag})_{s_1,s_1}\! -\! (\Lcal^{\dag})_{s_1,s_0}}\,.
\end{align}
}
Intuitively, we can view this expression as the influence of \rvs{node $s_1$ to node $s_0$}, normalized by the \rvs{sum of their mutual influence}. We quantify this influence with the following definition.  
\begin{definition}
\label{DomScore.def}
In a \rvs{strongly connected directed} graph $\Gcal$, the \emph{domination score} of node $u$ over $v$ is defined as
\rvs{
\begin{align}
D^{\Gcal}_{u,v} =(\Lcal^\dag)_{v,v} - (\Lcal^\dag)_{v,u}\,.
\end{align}
}
\end{definition}
\rvs{We provide two physical interpretations for $D^{\Gcal}_{u,v}$ in special cases. The first interpretation is that in a balanced regular (directed or undirected) graph, $D^{\Gcal}_{u,v}$ is the hitting time $H^{\Gcal}_{u,v}$.  A larger $H^{\Gcal}_{u,v}$ indicates that a random walker, starting from node $u$, spends more time in the network before it reaches $v$, therefore, exerting greater influence in the network. The second interpretation is that in an undirected graph $\Gcal$ and its induced electrical network $\overline{\Gcal}$, $\frac{1}{2m} D^{\Gcal}_{u,v}$ is the average voltage value of all nodes in the electrical network when unit current is injected at $u$ and extracted from $v$, and $v$ is grounded ($s_0$ has voltage $0$).} 

From the definition of domination score 
and the expression of \rvs{commute time} in Lemma~\ref{commute.def}, 
we immediately obtain
\begin{align}
\mu(S_1) = \frac{D^{\Gcal}_{s_1, s_0}}{C^{\Gcal}_{s_0,s_1}}= \frac{D^{\Gcal}_{s_1, s_0}}{D^{\Gcal}_{s_0,s_1}+ D^{\Gcal}_{s_1,s_0}}\,.
\end{align}
As for the deviation of the average opinion from the given value $\alpha$, we give its expression the following theorem.
\begin{theorem}
\label{compBalance.thm}
For \rvs{absolute} leader systems, if $|S_0|=|S_1|=1$,
\rvs{
\begin{align}
\label{compBalance.eqn}
f(S_1, \alpha) =  \frac{\left|(1-\alpha)D^{\Gcal}_{s_1, s_0} \!\!-\! \alpha D^{\Gcal}_{s_0,s_1}\right|}{D^{\Gcal}_{s_0,s_1}+ D^{\Gcal}_{s_1,s_0}}\,.
\end{align}
}
\end{theorem}
The proof of Theorem ~\ref{compBalance.thm} follows directly from (\ref{muSingleStubborn.eqn}) and Definition~\ref{DomScore.def}. 
 The numerator is the absolute value of a weighted average of $D^{\Gcal}_{s_1, s_0}$ and $-D^{\Gcal}_{s_0,s_1}$. 
Therefore, Theorem \ref{compBalance.thm} shows a weighted balance between the domination score of $s_0$ over $s_1$ and the domination score of $s_1$ over $s_0$, which decides the deviation of the average opinion from $\alpha$. 
Theorem \ref{compBalance.thm} indicates that for Problem~\ref{stubborn.prob}, if $|S_0|=|S_1|=1$, given the leader $s_0$, it suffices to find a node $s_1$ such that $(1-\alpha)D^{\Gcal}_{s_1, s_0} = \alpha D^{\Gcal}_{s_0,s_1}$ to shift the average opinion to $\alpha$. 

For \rvs{influenced} leader systems, the vector $\hat{\xx}$ is given by (\ref{partialSteady.eqn}). We do not apply the \rvs{leader-equivalent} graph analysis in this case  because $\Gcal' \neq \Gcal$.  We instead interpret $\hat{\xx}$ using properties of $\Gcal$.  Fortunately, when we choose one leader for each party, $\EE^{S_1}$ is a rank-$1$ matrix, and $\EE^{S} = \EE^{S_0} + \EE^{S_1}$ is a rank-$2$ matrix. Applying the rank-$1$ update of matrices twice leads to the following theorem.
\rvs{
\begin{theorem}
\label{partialBalance.thm}
For \rvs{influenced} leader systems, if $|S_0|=|S_1|=1$, we obtain 
\begin{align*}
f(S_1, \alpha)=\frac{\left|(1-\alpha)(\frac{\dd_{s_0}}{\kappa_0\ppi_{s_0}}+ D^\Gcal_{s_1,s_0})-\alpha(\frac{\dd_{s_1}}{\kappa_1\ppi_{s_1}}+ D^\Gcal_{s_0,s_1})\right|}{(\frac{\dd_{s_0}}{\kappa_0\ppi_{s_0}}+ D^\Gcal_{s_1,s_0}) + (\frac{\dd_{s_1}}{\kappa_1\ppi_{s_1}}+ D^\Gcal_{s_0,s_1})}\,,
\end{align*}
where the entries of the vector $\dd$ are defined as $\dd_v = \DD_{v,v}$, $\forall v\in\Vcal$.
\end{theorem}
We defer the proof of Theorem~\ref{partialBalance.thm} to Appendix~\ref{appendix4.sec}.
}

As observed in Theorem~\ref{compBalance.thm} for \rvs{absolute} leader systems, for \rvs{influenced} leader systems, Theorem \ref{partialBalance.thm} also shows the balancing behavior of domination scores in the social network, which decides the deviation of the average opinion from $\alpha$.
In addition, Theorem~\ref{partialBalance.thm} indicates that for Problem~\ref{partial.prob}, if $|S_0| = |S_1| = 1$, given the leader $s_0$, it suffices to find a node $s_1$ such that \rvs{$(1-\alpha)(\frac{\dd_{s_0}}{\kappa_{s_0}\ppi_{s_0}} + D^\Gcal_{s_1,s_0}) = \alpha(\frac{\dd_{s_1}}{\kappa_{s_1}\ppi_{s_1}} + D^\Gcal_{s_0,s_1})$ 
to shift the average opinion to $\alpha$.
} Assuming $\kappa_1 = \kappa_2$, \rvs{and they both approach infinity}, then the condition is the same as what we have derived in the \rvs{absolute} leader system.

The balancing behaviors shown in Theorem \ref{compBalance.thm} and \ref{partialBalance.thm} exhibit interesting results when \rvs{$\Gcal$ is undirected and} $\alpha = 1/2$. 
In particular, Theorems \ref{compBalance.thm} and \ref{partialBalance.thm} imply the following corollaries.
\begin{corollary}
\label{compBalance.cly}
For \rvs{absolute} leader systems,  \rvs{when $\Gcal$ is undirected}, $\alpha = 1/2$, and $|S_0|=|S_1|=1$,
\begin{align}
f(S_1, 1/2) = \frac{\left|\theta^{\Gcal}(s_0)^{-1} - \theta^{\Gcal}(s_1)^{-1}\right|}{2R^{\Gcal}_{s_0, s_1}}\,.
\end{align}
\end{corollary}

\begin{corollary}
\label{partialBalance.cly}
For \rvs{influenced} leader systems,  \rvs{when $\Gcal$ is undirected}, $\alpha = 1/2$, and $|S_0|=|S_1|=1$,
\begin{align}
f(S_1, 1/2)\! = \!\frac{\left|\theta^{\Gcal}(s_0)^{-1}\! +1/\kappa_0 - \theta^{\Gcal}(s_1)^{-1}\!-1/\kappa_1\right|}{2(R^{\Gcal}_{s_0,s_1}+1/\kappa_0 + 1/\kappa_1)}\,.
\end{align}
\end{corollary} 
These corollaries show the role of information centrality of leader nodes \rvs{in an undirected network} when the objective is to balance the opinions in the network. If $s_1$ has the same information centrality as $s_0$ (assuming $\kappa_1 = \kappa_0$ for \rvs{influenced} leader systems), then $\mu(S_1)= \frac{1}{2}$, and so the opinion network is balanced. If there is no such an $s_1$, then it is beneficial to find a node $s_1$ such that $|\theta^{\Gcal}(s_1)-\theta^{\Gcal}(s_0)|$ is small while $R^{\Gcal}_{s_0,s_1}$ is relatively large.

\subsection{Hardness of Choosing Optimal $k$ Leaders} 

\rvs{Next we show that} Problem~\ref{stubborn.prob} \rvs{is $\mathbf{NP}$-hard}. The hardness of Problem~\ref{partial.prob} remains an open question.

\begin{theorem}
\label{hardnessP1.thm}
The \rvs{\CSL} Selection  problem for shifting social opinion, described in Problem \ref{stubborn.prob}, is \textbf{NP}-hard.
\end{theorem}

\rvs{The proof of Theorem~\ref{hardnessP1.thm} is given in Appendix~\ref{appendix3.sec}.}

We note that in both Problems~\ref{stubborn.prob} and \ref{partial.prob}, $\mu(S_1)$, as a function of $S_1$, is monotone and submodular.
\begin{theorem}
\label{functionMnS.thm}
For both \rvs{absolute} and \rvs{influenced} leader systems, the set function $\mu(S_1)$ is monotone and submodular.
\end{theorem}
The monotonicity and submodularity of $\mu(S_1)$ for \rvs{influenced} leader systems follows  in a straightforward manner from results in~\cite{HZ18, MA19}. 
We are unaware of prior analogous results for \rvs{absolute leader} systems. We give simple proofs for both cases in Appendix~\ref{appendix7.sec}. Our proofs are based on analyzing the escape probabilities of random walks in the network.

\section{Algorithm}
\label{algorithm.sec}

In this section, we present an algorithm for selecting a set of nodes to be leaders in $S_1$,  given set of leaders $S_0$, to shift the average opinion as close as possible to a given value $\alpha$. 

\rvs{\subsection{Algorithm Intuition}
}

\rvs{
It is well known that greedy algorithms give a $(1-1/e)$ approximation for monotone submodular maximization problems with cardinality constraints~\cite{NWF78}.  According to Theorem~\ref{functionMnS.thm}, for either Problem \ref{stubborn.prob} or \ref{partial.prob}, a greedy algorithm provides a $(1-1/e)$ approximation for the problem when $\alpha = 1$. However, for other values of $\alpha$, the problems are not trivial to solve. We observe that if $\mu(S_1)\leq \alpha$ always holds, we have a submodular maximization problem with cardinality constraint; if $\mu(S_1)\geq \alpha$ always holds, the problem  is a submodular minimization problem with the same cardinality constraint. However, we do not know the value of $\mu(S_1)$ beforehand. Therefore, we need to design a more sophisticated algorithm to approximately solve Problem \ref{stubborn.prob} and \ref{partial.prob}.
}

The intuition behind our algorithm is to consider these problems as  submodular cost
submodular knapsack (SCSK) constraint maximization problems~\cite{AN09,IB13}.  
An SCSK constrained maximization problem is defined as $$\textrm{maximize } f(X) \quad \textrm{subject to } g(X)\leq b.$$
for submodular functions $f$ and $g$, and upper bound $b\in \mathbb{R}$.
Problems \ref{stubborn.prob} and \ref{partial.prob} can be interpreted as special cases of SCSK with additional cardinality constraints: 
\begin{align}
\label{SCSKCard.prob}
\textrm{maximize  }& \mu(S_1) \,\nonumber\\
 \textrm{subject to:  }& S_1\subseteq Q, \mu(S_1)\leq b, |S_1|\leq k.
\end{align}

Our algorithm is motivated by an approach in~\cite{IB13} for the general SCSK problem. We approximate the optimum $\mu(S_1)$ for Problem \ref{stubborn.prob} or \ref{partial.prob} by imposing an upper bound for the submodular function $\mu$ and then applying a submodular maximization algorithm to the bounded problem. Specifically, we find an appropriate upper bound constraint $\mu(S_1)\leq b$, such that a greedy algorithm for maximizing $\mu(S_1)$ with upper bound $b$ leads to an approximation algorithm for optimum solution $S^{*}$, where
$$S^* \in \argmin_{P\subseteq Q, |P|\leq k} |\mu(P)-\alpha|$$
is an optimal solution for Problem \ref{stubborn.prob} or \ref{partial.prob}, respectively.

We apply a greedy algorithm to problem (\ref{SCSKCard.prob}).  For an upper bound $b$, the algorithm $\rm{Greedy}$ returns a solution $S_b$. We can compare different upper bounds by the solutions $\rm{Greedy}$ returns. The bound $b_1$ is a better upper bound than $b_2$ if $|\mu(S_{b_1})-\alpha| < |\mu(S_{b_2})-\alpha|$. We further define the best upper bound  input for algorithm $\rm{Greedy}$ as $b^*$, or formally,
\begin{align}
\label{bestBound.def}
b^* \in \argmin_{b\in[\alpha, 1]} |\mu(S_{b})-\alpha|\,.
\end{align}
We use a modified binary search to converge to the best upper bound $b^*$ for $\rm{Greedy}$. 
In the next subsection, we describe both the bound search algorithm and the routine $\rm{Greedy}$.

\subsection{Bounded Search Approximation Algorithm}
We first define the algorithm in terms of Problem~\ref{stubborn.prob}.  We describe the changes of the algorithm in order to solve Problem~\ref{partial.prob} in the end of the subsection.

\begin{algorithm}
    \caption{$P = \textrm{BoundSearch}(\Gcal, Q, \alpha, k, \delta)$}
	\label{alg:boundSearch}
    $\svs \gets \emptyset$\;
    $b_{\min} \gets \alpha$; $b_{\max} \gets 1$\;
    $\hat{b} \gets 1$ \qquad \tcp{current upper bound}
    $\check{b}\gets \hat{b}$ \qquad \tcp{current best upper bound}
    $d_{\min} \gets \alpha$ \qquad \tcp{minimal $|\mu - \alpha|$ so far}
    $t \gets 0$\;
    \Do{$\frac{b_{\max}}{b_{\min}} > \exp(\delta)$}{
    		$t \gets t+1$\;
    		$(S, \mu) = \textrm{Greedy}( \Gcal, Q, \hat{b}, k)$\;
    		$\hat{d} \gets |\mu-\alpha|$\;
    		\If{$\hat{d} < d_{\min}$}{
    				$P \gets S$; $d_{\min} \gets \hat{d}$; $\check{b} \gets \hat{b}$
    		}
    		\eIf{$\mu > \alpha$}{
    				\While{ $\mu \leq (b_{\min}+b_{\max})/2$}{
    						$b_{\max} \gets (b_{\min}+b_{\max})/2$
    				}
    		}{
    			$b_{\min} \gets \hat{b}$\;
    			\If{$(\alpha + \hat{d}) < b_{\max}$}{
    				$b_{\max} \gets (\alpha + \hat{d})$\;
    				$\hat{b} \gets b_{\max}$; \Continue
    			}
    		}
    		$\hat{b} \gets (b_{\min}+b_{\max})/2$
    }
    \KwRet{\svs}
\end{algorithm}

Our algorithm, \textrm{BoundSearch}, is given in Algorithm~\ref{alg:boundSearch}.
The algorithm takes  as input a graph $\Gcal$, a candidate vertex set $Q$,  
an objective opinion $\alpha$,  a cardinality constraint $k$, and a precision parameter $\delta$ for binary search.  It returns a set of nodes $P$, which is a subset of $Q$ satisfying $|P|\leq k$.

The bound $\hat{b}$ is initialized with value $1$, and the algorithm searches for $b^*$ in the interval $[b_{\min}, b_{\max}]$ that 
might include a better upper bound than $\check{b}$, the current best bound found by the algorithm that leads to the smallest $|\mu(S_b)-\alpha|$. 
We update $b_{\min}$ and $b_{\max}$ until $b_{\min}\Approx{\delta}b_{\max}$, and $b^*,\check{b},\hat{b}\in [b_{\min}, b_{\max}]$. We obtain $\check{b}\Approx{\delta}b^*$. Since $\check{b}$ is the current best upper bound found by the algorithm, for any $b\notin [b_{\min}, b_{\max}]$, $|S_{\check{b}}-\alpha| \leq |S_{b}-\alpha|$.

Before analyzing Algorithm~\ref{alg:boundSearch}, we recall the concept of $\eps$-approximation~\cite{PS14}:
\begin{definition}
Given two numbers $a, b\in \mathbb{R}$, $a,b \geq 0$, if
$$\exp(-\eps)a \leq b \leq \exp(\eps)a\,,$$
then $a$ is an \emph{$\eps$-approximation} of $b$, denoted by $a \Approx{\eps} b$.
\end{definition}
Note that $a \Approx{\eps} b$ if and only if $b \Approx{\eps} a$.

In Algorithm~\ref{alg:Greedy}, we present the greedy routine $P = \textrm{Greedy}(\Gcal, Q, \hat{b}, k)$ for the constrained submodular maximization described in (\ref{SCSKCard.prob}). The algorithm takes as input a graph $\Gcal$, a candidate set $Q$, an SCSK upper bound $\hat{b}$, and an integer $k$ for the cardinality constraint.  It returns a set of nodes $P$, which is a subset of $Q$ satisfying $|P|\leq k$ and $\mu(P)\leq \hat{b}$. The algorithm chooses the node that most increases $\mu(P)$ without violating the upper bound from the candidate set in each iteration, deletes it from the candidate set, and adds it to the current leader set.

\begin{algorithm}[htbp]
    \caption{$(P,\mu) = \textrm{Greedy}(\Gcal, Q, \hat{b}, k)$}
	\label{alg:Greedy}
    $\svs \gets \emptyset$\;
    \While{$|Q|>0$ \bf{and} $|P| < k$}{
    		$s \gets \argmax_{u\in Q} \mu(P\cup \{u\})$\;
    		\If{$\mu(P\cup\{s\})\leq \hat{b}$}{
    				$P \gets (P\cup\{s\})$
    		}
    		$Q \gets (Q\backslash \{s\})$
    }
    $\gamma \gets \mu(P)$\;
    \KwRet{$(\svs, \gamma)$}
\end{algorithm}


To analyze Algorithm~\ref{alg:Greedy}, we introduce the concept of the minimum cover number.
\begin{definition}
The \emph{minimum cover number} $k_{\mu,b}$ for set function $\mu(S)$, $S\subseteq Q$, and $b\in\mathbb{R}$ is defined as
\begin{align*}
k_{\mu,b} \defeq \min\{|S|: \mu(S)\geq b\}\,,
\end{align*}
if there exists $S$ satisfying $\mu(S)\geq b$, otherwise $k_{\mu,b} \defeq +\infty$. 
\end{definition}


%
%
%
%


%
Then, we obtain the approximation ratio of $\rm{BoundSearch}$.
\begin{theorem}
\label{algBound.thm}
Consider a graph $\Gcal$, a candidate set $Q$, 
an objective $\alpha$, a cardinality constraint $|P|\leq k$, and a precision parameter $\delta>0$.
Let $S^*$ be an optimal solution for Problem~\ref{stubborn.prob} for these parameters. 
The algorithm $P ={\rm{BoundSearch}}(\Gcal, Q, \alpha, k, \delta)$ returns a node set $P$ such that $\mu(P)\Approx{\sigma}\mu(S^*)\,,$
in which $\sigma = -\ln(1-\zeta)+\delta$, and $\zeta\defeq \max(1/e, 1/k_{\mu,\alpha})$. 
\end{theorem}

\rvs{We defer the proof of Theorem~\ref{algBound.thm} to Appendix~\ref{appendix5.sec}.}

The guarantee given in Theorem~\ref{algBound.thm} can also be written as:
\begin{align*}
\kh{1-\zeta} e^{-\delta} \mu(S^*) \leq \mu(P) \leq \kh{1-\zeta}^{-1} e^\delta \mu(S^*)\,.
\end{align*}

The ${\rm{BoundSearch}}$ algorithm can be applied to Problem~\ref{partial.prob} with the same approximation guarantee with the only difference that the stubbornness function $\kappa$ is an input of the algorithm. The stubbornness function is also passed into ${\rm{Greedy}}$ to calculate the average opinion. Theorem~\ref{algBound.thm} holds for the corresponding algorithm $P ={\rm{BoundSearch}}(\Gcal, Q, \alpha, k, \kappa,\delta)$, which calls ${\rm{Greedy}}(\Gcal, Q, \hat{b}, k,\kappa)$.

\subsection{Complexity Analysis}
A naive implementation of the proposed algorithm runs in $O(kqn^3\log{\frac{1}{\delta}})$ time, which is expensive for large graphs. Using blockwise inversion and rank-$1$ update of matrices we can improve the running time of  $\rm{BoundSearch}$ to $O(n^3\log{\frac{1}{\delta}})$.

\begin{theorem}
\label{complexity.thm}
There exists an implementation of Algorithm~\ref{alg:boundSearch} for a graph with $n$ nodes that has running time $O(n^3\log{\frac{1}{\delta}})$.
\end{theorem}
\rvs{The proof of Theorem~\ref{complexity.thm} is given in Appendix~\ref{appendix0.sec}.}



\section{Experiments}
\label{experiments.sec}

In this section, we present experiments to highlight the analytical results and to show the effectiveness of the proposed algorithm. 
\begin{table}
    \centering
    \newcolumntype{P}[1]{>{\centering\arraybackslash}p{#1}}
    \fontsize{8}{10}\selectfont 
    \scalebox{.8}{\rvs{%
    \begin{tabular}{||P{0.75cm} || P{1.5cm} | P{1.5cm} | P{1.5cm} | P{1.5cm} | P{1.5cm} | P{1.5cm}||}
        \hline 
         $\alpha$ & Optimum & 
         DS & ER & Random \\
        \hline
        \hline
0.25 & 0.249830 & 0.250214 & 0.000083 & 0.248751 \\ 
0.50 & 0.499699 & 0.500495 & 0.000083 & 0.122848  \\
0.75 & 0.750014 & 0.750014 & 0.000083 & 0.002391  \\
1.00 & 0.999645 & 0.999645 & 0.000083 & 0.576522  \\
        \hline
    \end{tabular}
    }
    }
    \caption{Average opinion in an \rvs{absolute} leader system. The graph is 
        \rvs{the Twitter Retweet Network \emph{rt-higgs}} with a fixed $s_0$ and a 
        node $s_1$ chosen via various methods. 
    }
    \label{singleLeader.table}
\end{table}

\begin{table}
    \centering
    \newcolumntype{P}[1]{>{\centering\arraybackslash}p{#1}}
    \fontsize{8}{10}\selectfont 
    \scalebox{.8}{\rvs{%
    \begin{tabular}{||P{0.75cm} || P{1.5cm} | P{1.5cm} | P{1.5cm} | P{1.5cm} | P{1.5cm} | P{1.5cm}||}
        \hline 
         $\alpha$ & Optimum & 
         DS\&K & ER & Random \\
        \hline
        \hline
0.25  & 0.250010 & 0.250010 & 0.000028 & 0.327045 \\
0.50  & 0.500010 & 0.500010 & 0.000028 & 0.345662 \\
0.75  & 0.750111 & 0.750111 & 0.000028 & 0.305945 \\
1.00  & 0.997124 & 0.997124 & 0.000028 & 0.000000 \\
        \hline
    \end{tabular}
    }
    }
    \caption{Average opinion in an \rvs{influenced} leader system. The graph is 
        \rvs{the Twitter Retweet Network \emph{rt-higgs}} with a fixed $s_0$ and a 
        node $s_1$ chosen via various methods. 
    }
    \label{singleLeader2.table}
\end{table}


We first study the properties of $\mu(S_1)$ when $|S_0|=|S_1|=1$ \rvs{in absolute and influenced leader systems} for $\alpha = 0.25, 0.5, 0.75,$ and $1$. 
The leader $s_0$ is chosen uniformly at random.
\rvs{We run experiments on a directed and
weighted social network. We utilize the largest strongly connected component
of the Twitter Retweet Network with the keyword ``higgs",
which we refer to as \emph{rt-higgs}~\cite{nr}. The edges are weighted 
by the number of retweets to a user.
The network has $13,086$ nodes and $63,537$ edges.}

\rvs{For the absolute leader system}, 
we find the average opinion of the network for the optimal solution to 
Problem~\ref{stubborn.prob}, i.e., the optimal $s_1$ as given by 
Theorem~\ref{compBalance.thm}. We also show the 
average opinion when $s_1$ is chosen using heuristics motivated by the theorem.  
The first heuristic, DS, is based on the domination score; we find 
the $s_1$ such that the resulting $\mu(\{s_1\})$ minimizes the 
numerator of~(\ref{compBalance.eqn}).
We also use a heuristic based on effective resistance (ER); 
here, $s_1$ is chosen so as to maximize the 
denominator of~(\ref{compBalance.eqn}).
Finally, we compute the average opinion for a randomly 
chosen $s_1$. The results of this experiment are shown in Table~\ref{singleLeader.table},

We also conduct an experiment for \rvs{an influenced} leader system
using the \rvs{\emph{rt-higgs}} network. \rvs{Influenced} leaders have 
uniform stubbornness $\kappa=1$, and the other parameters 
are the same as the experiment for \rvs{the \csl} system. 
We find the optimal $s_1$ as well as the $s_1$ chosen by 
heuristics motivated by the numerator (DS\&K) and denominator (ER) 
of the result given in Theorem \ref{partialBalance.thm}. 
We note that the $s_1$ that minimizes denominator of the 
result in Theorem~\ref{compBalance.thm} also minimizes 
denominator of the result in Theorem~\ref{partialBalance.thm}. 
The results are shown in Table~\ref{singleLeader2.table}.  

Table~\ref{singleLeader.table} and \ref{singleLeader2.table} show 
that when $|S_0|=|S_1|=1$,  the domination score well captures the 
behavior of $\mu(\{s_1\})$. We have observed similar results in 
various  Erd\H{o}s–R\'{e}nyi graphs with different choices of a 
single leader $s_0$. 

%
%

\begin{figure}[htbp]
\centering
{
\begin{subfigure}{.24\textwidth}
    \centering
    \includegraphics[scale=0.24]{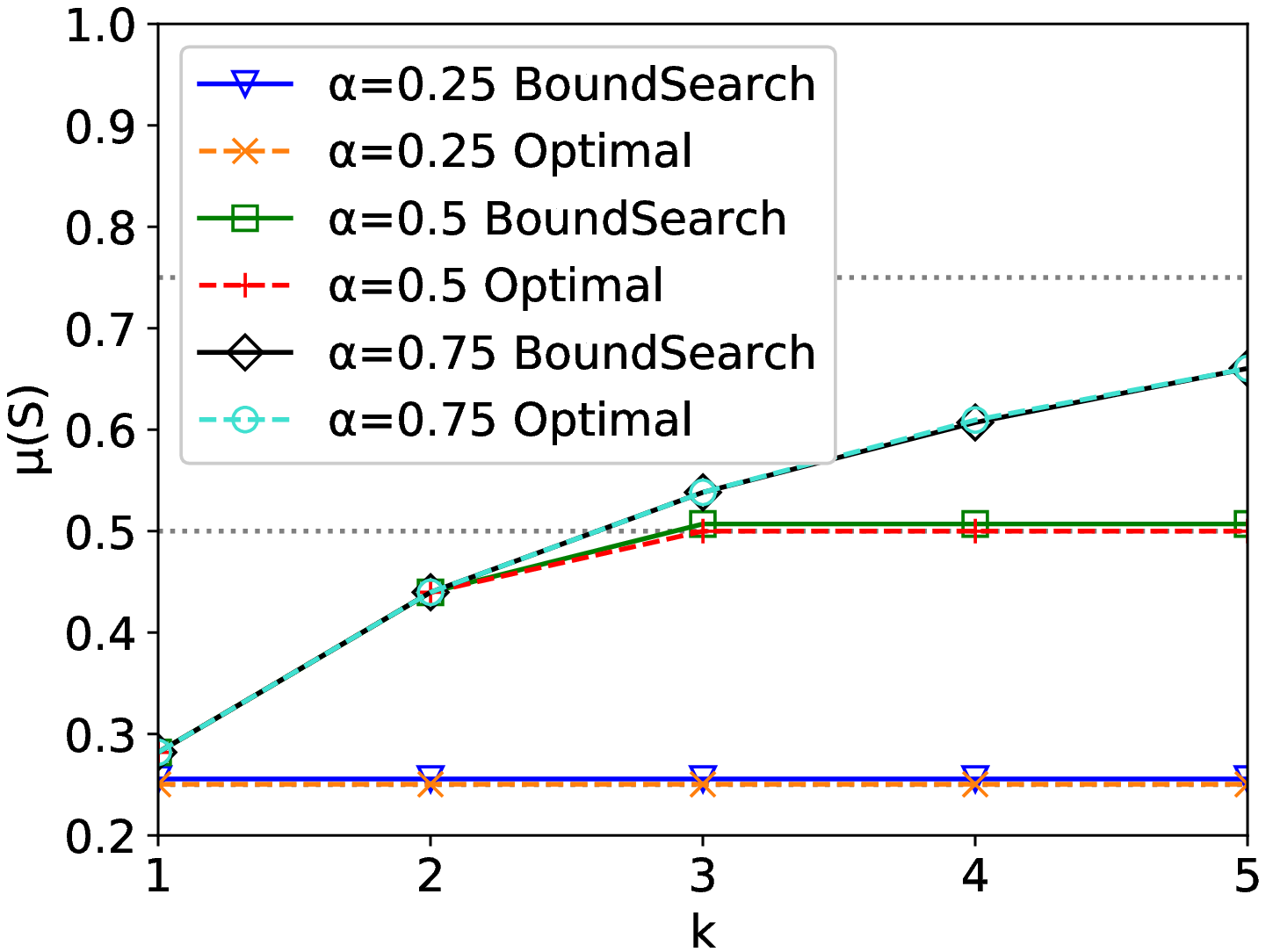}
    \caption{\rvs{Absolute leader system.}}
    \label{exp2_fig}
\end{subfigure}
\begin{subfigure}{.24\textwidth}
    \includegraphics[scale=0.24]{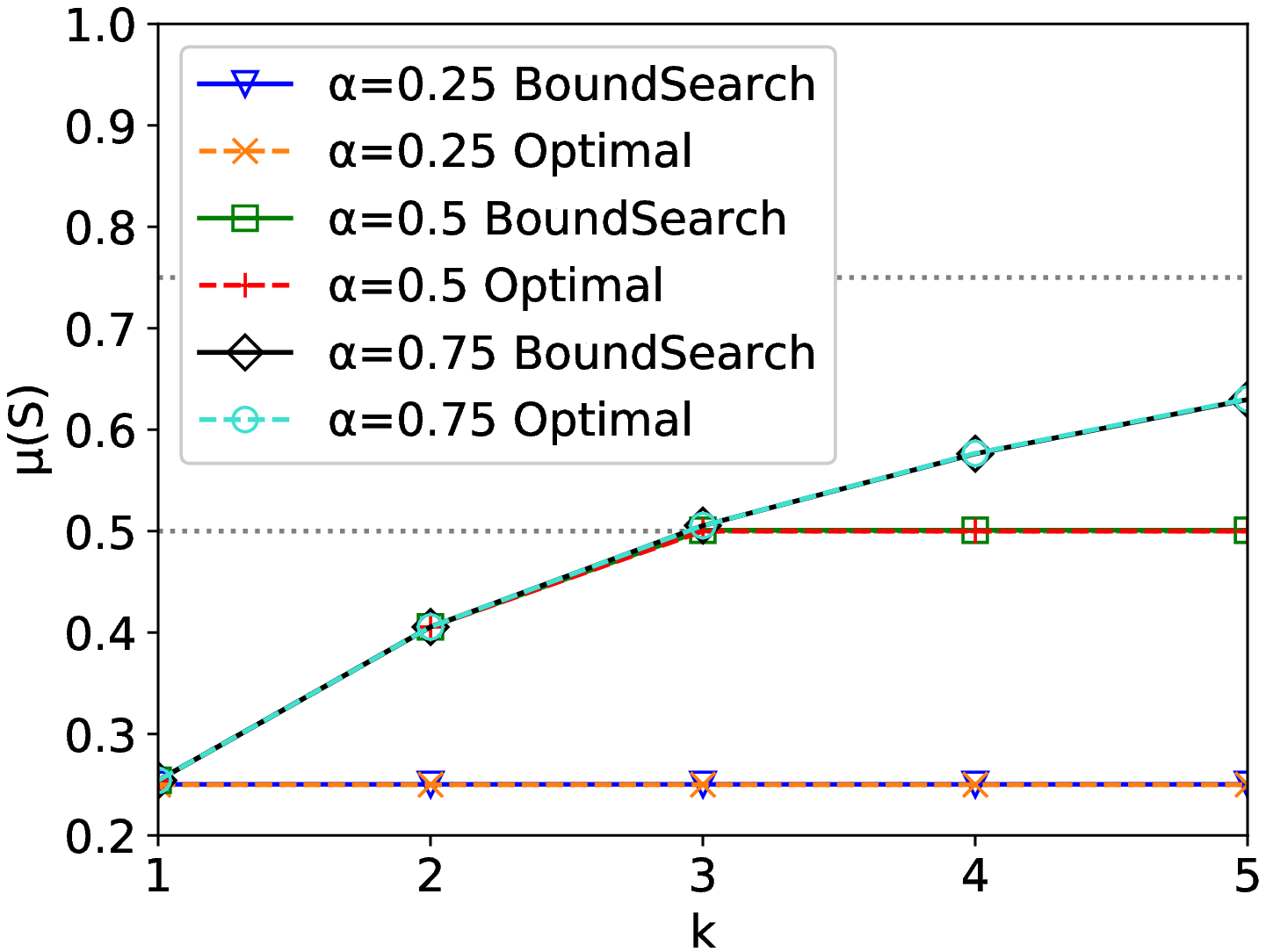}
    \caption{\rvs{Influenced leader system.}}
    \label{exp2_partial_fig}
\end{subfigure}
}
\caption{Average opinion of Optimum vs. average opinion of \rm{BoundSearch} in an Erd\H{o}s–R\'{e}nyi graph with $30$ nodes  and connecting probability $0.1$. $S_0$ leader sets are chosen randomly, and $S_1$ leader sets are chosen by brute-force search and \rm{BoundSearch} with different $k$ and $\alpha$ values.}
\label{algVSOpt.fig}
\end{figure}

Next, to show the effectiveness of our leader selection algorithm, 
we compare the result returned by our algorithm \rm{BoundSearch} 
with the optimal value returned by brute-force search. 
We use an \rvs{unweighted undirected} Erd\H{o}s–R\'{e}nyi graph with $30$ nodes and connecting probability $0.1$. 
We choose an $S_0$ leader set  of size $3$ at random. We run the \rm{BoundSearch} algorithm for both \rvs{absolute} leader system and \rvs{influenced} leader systems with $\alpha \in \{0.25, 0.50, 0.75\}$. \rvs{Influenced} leaders use uniform stubbornness $\kappa =1$. 
 The results are shown in Figure~\ref{algVSOpt.fig}.  In all cases, \rm{BoundSearch} returns nearly optimal results.

\begin{table}[htbp]
    \centering
    \newcolumntype{P}[1]{>{\centering\arraybackslash}p{#1}}
    \fontsize{8}{10}\selectfont 
    \scalebox{.8}{\rvs{%
    \begin{tabular}{||P{0.75cm} || P{1.5cm} | P{1.5cm} | P{1.5cm} | P{1.5cm} | P{1.5cm} ||}
        \hline 
        $\alpha$ & \rm{BoundSearch} & PDS & Random \\
        \hline
        \hline
0.25 & 0.250730 & 0.237610 & 0.428106 \\
0.50 & 0.500975 & 0.565233 & 0.206765 \\
0.75 & 0.750976 & 0.843537 & 0.495980 \\
1.00 & 1.000000 & 1.000000 & 0.466443 \\
        \hline
    \end{tabular}
    }
    }
\caption{Average opinion in an internal leader system 
    on the \emph{rt-higgs} network with $|S_0| =100$ and $k=100$.}
\label{exp5.table1}
\end{table}
\begin{table}[htbp]
    \centering
    \newcolumntype{P}[1]{>{\centering\arraybackslash}p{#1}}
    \fontsize{8}{10}\selectfont 
    \scalebox{.8}{\rvs{%
    \begin{tabular}{||P{0.75cm} || P{1.5cm} | P{1.5cm} | P{1.5cm} | P{1.5cm} | P{1.5cm} ||}
        \hline 
        $\alpha$ & \rm{BoundSearch} & PDS\&K & Random \\
        \hline
        \hline
0.25  & 0.250732 & 0.252612 & 0.640516\\
0.50  & 0.500976 & 0.499114 & 0.418073\\
0.75  & 0.750976 & 0.748867 & 0.455767\\
1.00  & 0.975863 & 0.973205 & 0.533567\\
        \hline
    \end{tabular}
    }
    }
\caption{Average opinion in an external leader system 
    on the \emph{rt-higgs} network with $|S_0| =100$ and $k=100$.}
\label{exp5.table2}
\end{table}

\rvs{We next run our leader selection algorithm 
on \emph{rt-higgs} with $|S_0|  = 100$ and $k = 100$.
We compare 
the result produced by our algorithm \rm{BoundSearch} with a heuristic we call the
Propositional Domination Score and with a randomly select set. 
To calculate the Proposition Domination Score, 
for each leader in $S_0$, we choose a leader 
for $S_1$  according to 
Theorem~\ref{compBalance.thm} for absolute leaders
and~\ref{partialBalance.thm} for influenced leaders. 
For all influenced leaders, $\kappa=1$.
Tables \ref{exp5.table1} and \ref{exp5.table2} 
shows that our algorithm converges to the desired value and outperforms
the heuristic in all tested cases.}

\begin{figure}[htbp]
\centering
\begin{subfigure}[b]{.24\textwidth}
    \centering
    \includegraphics[scale=0.24]{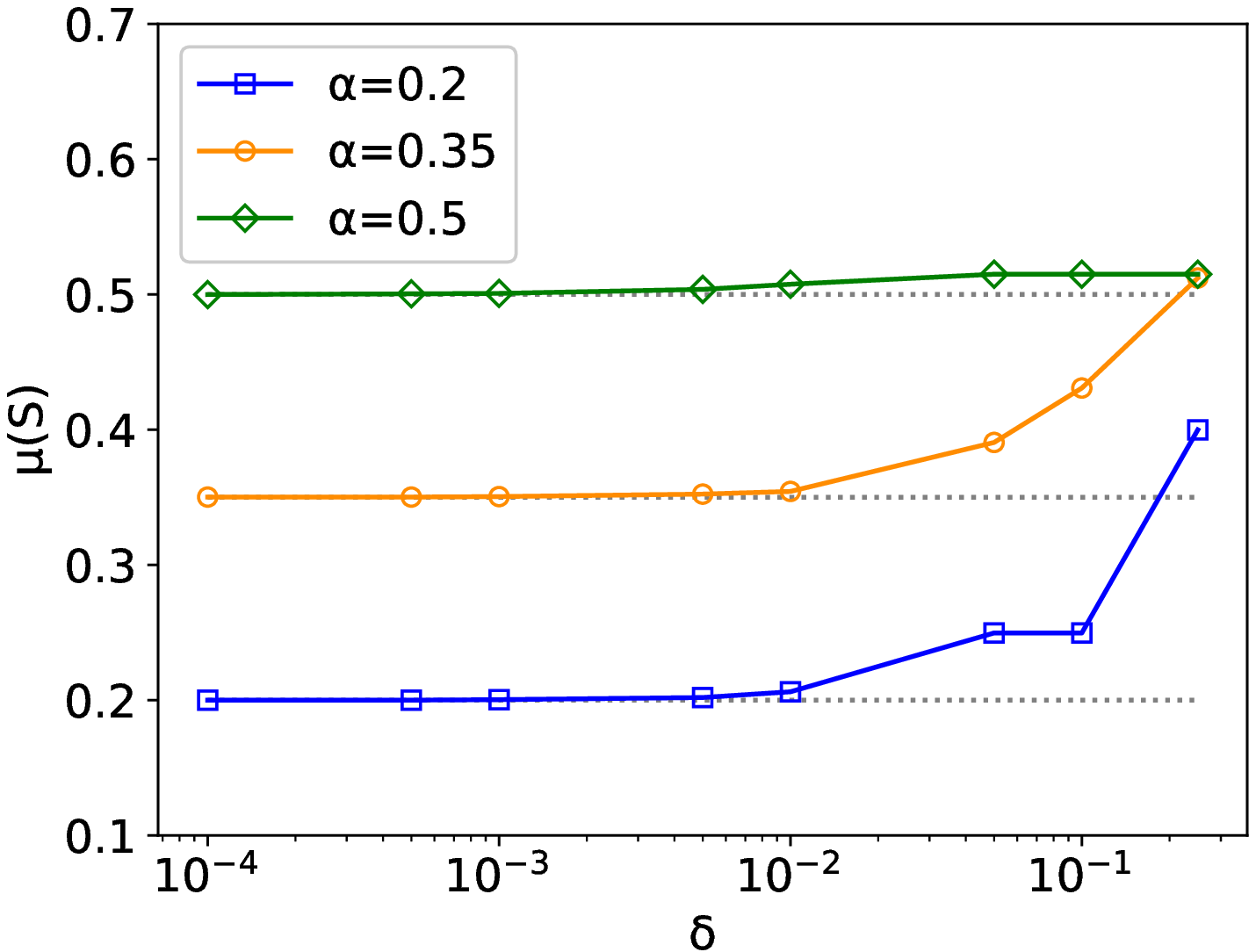}
    \caption{\rvs{Internal Leader System}}
    \label{exp3_fig}
\end{subfigure}
\begin{subfigure}[b]{.24\textwidth}
    \includegraphics[scale=0.24]{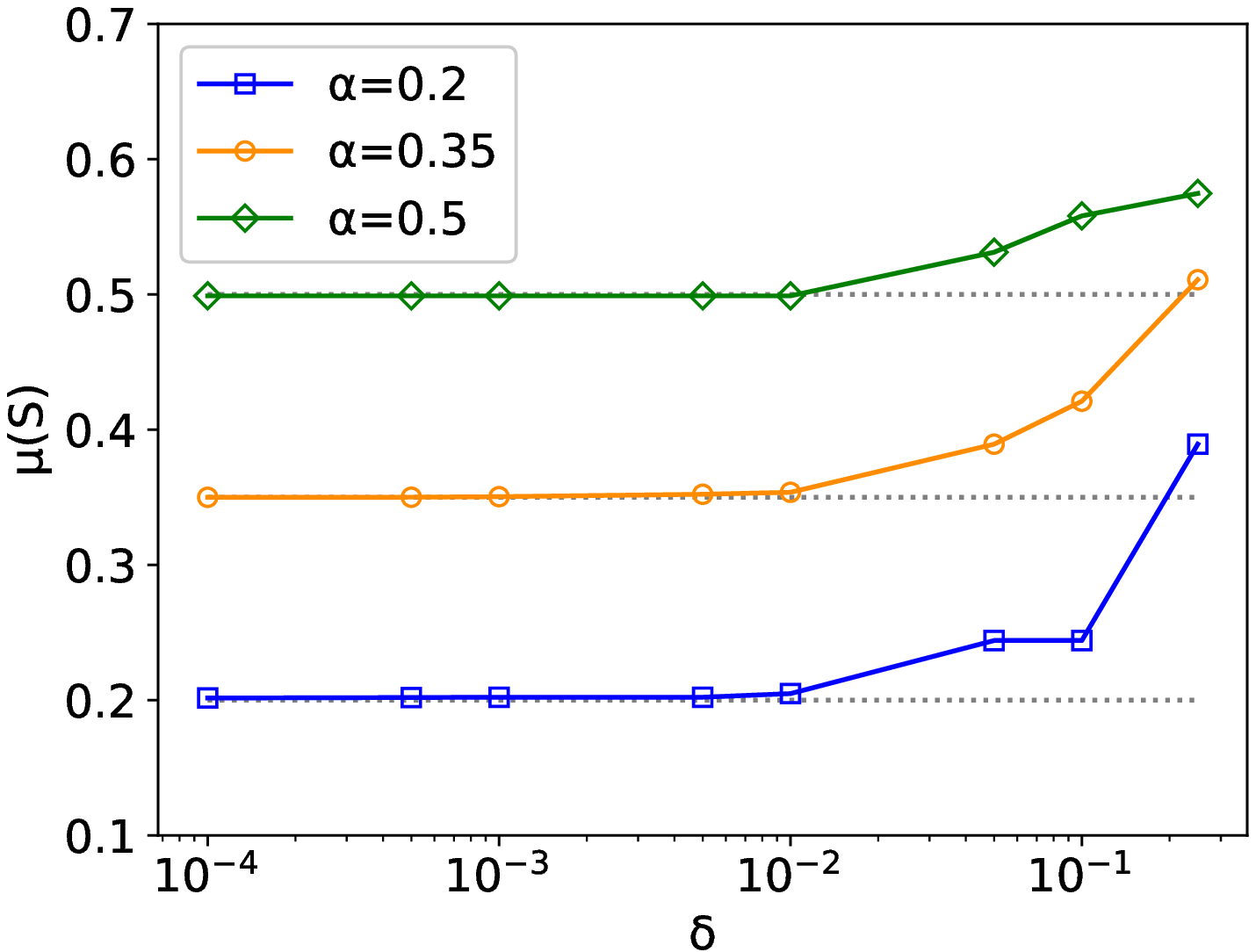}
    \caption{\rvs{External Leader System}}
    \label{exp4_partial_fig}
\end{subfigure}
\caption{Effect of varying $\delta$ on \rm{BoundSearch} at 
${\alpha \in \{0.20, 0.35, 0.50\}}$. Experiment uses the Haggle graph.}
\label{exp4.fig}
\end{figure}

Finally, we explore the effect of varying the $\delta$ 
parameter in  $\rm{BoundSearch}$. We run $\rm{BoundSearch}$
on the Haggle~\cite{CHC+07} social contact graph. The Haggle graph is a
multigraph, which we turn it into an \rvs{undirected} simple graph by deleting all duplicate edges.
We use the largest connected component of the graph which has $274$ nodes and $2124$ edges. \rvs{All edges have unit edge weight.}
We set $k=15$ and $\alpha \in \{0.2, 0.35, 0.5\}$. 
We vary $\delta$ from $0.0001$ to $0.25$. For the \rvs{absolute leader system}, we have $|S_0|=80$, and for the \rvs{influenced leader system},
leaders we have $|S_0|=15$. \rvs{Influenced} leaders use uniform stubbornness $\kappa =1$. The results are shown in Figure~\ref{exp4.fig}. 
We observe that as $\delta$ 
decreases, the results from \rm{BoundSearch} converge to
a value close to $\alpha$.

\section{Conclusion} \label{conclusion.sec}

We have  studied two French-DeGroot opinion dynamics models where leaders have polarizing opinions. 
For both models, we showed expressions for the steady-state opinion using the Laplacian matrix of a \rvs{leader-equivalent} graph. For the single leader case, we gave an explicit expression for the steady-state opinion vector and analyzed the average opinion based on the expression. 
Then, we studied the problem of shifting the average steady-state opinion to a given value by selecting an opposing leader set with a cardinality constraint. We gave both a hardness result for this problem and an algorithm with provable approximation ratio. 
We also presented experiments showing that our algorithm returns results close to optimal in practice. 
Future work will focus on algorithms with better approximation ratios and running time and the hardness of the \rvs{\psl} selection problem.




\appendix
\subsection{Some Useful Matrix Identities}
We introduce some matrix identities. 
\label{appendix1.sec}
\rvs{
\begin{lemma}
\label{MatrixEqn1}
For any $\pp\perp \one$, $\rr\perp\one$,
$$\pp^{\top} \Lcal^{\dag} \rr = \pp^{\top} \Rcal \rr\,.$$
\end{lemma}
The proof was given in \cite[Appendix C.2]{CKPP+16b}.

\begin{lemma}
\begin{align*}
\II - \LL\LL^{\dag} &= \frac{1}{\|\DD^{-1}\ppi\|^2} \DD^{-1}\ppi \ppi^{\top}\DD^{-1}\,,\\
\II - \LL^{\dag}\LL &= \frac{1}{n}\one\one^{\top}\,.
\end{align*}
\end{lemma}
\begin{IEEEproof}
\begin{align*}
\II - \LL\LL^{\dag}  &= \II - \II_{\Ima(\LL)} = \II_{\ker(\LL^\top)}\,,\\
\II - \LL^{\dag}\LL  &= \II - \II_{\Ima(\LL^\top)} = \II_{\ker(\LL)}\,,
\end{align*}
which completes the proof.
\end{IEEEproof}

\begin{lemma}
For any $\pp\perp \one $, $\rr\perp\one$, $$\pp^\top\LL^{\dag}\DD\PPi^{-1}\rr = \pp^\top\Lcal^{\dag}\rr\,.$$
\end{lemma}
\begin{IEEEproof}
From Lemma~\ref{MatrixEqn1} we know that $\pp^{\top}\Lcal^{\dag}\rr=\pp^{\top}\Rcal^{\dag}\rr=\pp^\top(\II-\WW^\top)^\dag\PPi^{-1}\rr$. Therefore it suffices to prove
\begin{align}
\label{condition.eqn}
\pp^{\top}\!((\DD(\II-\WW^\top))^\dag \DD\PPi^{-1}\! -(\II - \WW^\top)^\dag \PPi^{-1})\rr\! = \!0\,.
\end{align}
Since
\begin{align*}
&\pp^{\top}\kh{(\DD(\II-\WW^\top))^\dag \DD -(\II - \WW^\top)^\dag }(\II- \WW^\top)\\
=& \pp^{\top} \kh{\II_{\Ima(\II-\WW)\DD} - \II_{\Ima(\II-\WW)}}
\end{align*}
and $\pp\perp\one$, $\one\in\ker(\II-\WW^\top)$, $\one\in\ker(\DD(\II-\WW^\top))$, then we attain $\pp\in \Ima(\II-\WW)$ and $\pp\in \Ima((\II-\WW)\DD)$, which leads to
\begin{align*}
&\pp^{\top}\kh{(\DD(\II-\WW^\top))^\dag \DD -(\II - \WW^\top)^\dag }(\II- \WW^\top)=0\,.
\end{align*}
Therefore
\begin{align*}
\kh{(\DD(\II-\WW))^\dag \DD -(\II - \WW)^\dag }\pp \in \ker (\II - \WW)
\end{align*}
Then we know that
\begin{align*}
&\pp^{\top}\kh{(\DD(\II-\WW^\top))^\dag \DD -(\II - \WW^\top)^\dag }\PPi^{-1}\rr\\
= &\vv^{\top} \PPi^{-1}\rr =\rr^{\top}\PPi^{-1}\vv\,,
\end{align*}
where $\vv\in \ker(\II-\WW)$. Therefore $\rr^\top\PPi \vv = \rr^\top\PPi^{-1}\ppi \cdot \beta$, $\beta$ is a scaling factor. Since $\rr^\top\PPi^{-1}\ppi = \rr^{\top}\one =0$, we attain (\ref{condition.eqn}), which proves the lemma.
\end{IEEEproof}

\begin{lemma} 
\label{MatrixEqn4}
For any $\yy \perp \one$, 
$$\kh{ \PPi(\II - \WW^{\top})}\kh{(\II - \WW^{\top})^{\dag} \PPi^{-1}} \yy= (\II - \frac{1}{n}\one\one^{\top})\yy\,.$$
\end{lemma}
\begin{IEEEproof}
It suffices to prove that 
$$\kh{ \PPi(\II - \WW^{\top})}\kh{(\II - \WW^{\top})^{\dag} \PPi^{-1}} \yy= \yy \,.$$
Since $\PPi^{-1}\yy \perp \ppi$ and $(\II - \WW)\ppi = 0$, therefore $\PPi^{-1}\yy \in \Ima(\II - \WW^{\top})$. Then $\PPi(\II - \WW^{\top})(\II - \WW^{\top})^{\dag} (\PPi^{-1} \yy) = \PPi \II_{\Ima(\II - \WW^{\top})} (\PPi^{-1}\yy) = \PPi \PPi^{-1}\yy = \yy$.
\end{IEEEproof}
}
\subsection{Proof of Proposition \ref{explicitProb.thm}}
\label{appendix2.sec}
\rvs{
\begin{IEEEproof}
We can express (\ref{sys1Leaders.sys}) and (\ref{sys1Followers.sys}) in the following form
\begin{align*}
\begin{pmatrix}\dot{\xx}_{S}(t)\\
\dot{\xx}_{F}(t)\end{pmatrix}
=-\begin{pmatrix}
\zero & \zero\\
\LL_{F,S} & \LL_{F,F}
\end{pmatrix}
\begin{pmatrix}\xx_{S}(t)\\
\xx_{F}(t)\end{pmatrix}\,.
\end{align*}
When the equilibrium is reached,
\begin{align*}
\begin{pmatrix}
\zero & \zero\\
\LL_{F,S} & \LL_{F,F}
\end{pmatrix}
\begin{pmatrix}\hat{\xx}_{S}(t)\\
\hat{\xx}_{F}(t)\end{pmatrix} = \zero\,,
\end{align*}
Since $\LL =\DD (\II - \WW^{\top})$, this is equivalent to solving
\begin{align*}
\begin{pmatrix}
\zero & \zero\\
[\II - \WW^{\top}]_{F,S} & [\II - \WW^{\top}]_{F,F}
\end{pmatrix}
\begin{pmatrix}\hat{\xx}_{S}(t)\\
\hat{\xx}_{F}(t)\end{pmatrix} = \zero\,,
\end{align*}
When $S_0=\{s'_0\}$ and $S_1=\{s'_1\}$, $\xx_{S}(t) = (1 \,\,0)^\top$; it suffices to solve
\begin{small}
\begin{align}
\label{systemExpand}
\begin{pmatrix}
1  & 0 &  \zero\\
0 & 1 & \zero\\
[\II \!\!-\!\! \WW^{\top}\!]_{F\!,s'_0}\!\!\!\!\!\!\! & [\II \!\!-\!\! \WW^{\top}\!]_{F\!,s'_1}\!\!\!\!\!\!\! & [\II \!\!-\!\! \WW^{\top}\!]_{F\!,F}\!\!
\end{pmatrix}
\!\!\! \begin{pmatrix}0\\1\\
\hat{\xx}_{F}(t)\end{pmatrix} \!\!=\!\!
\begin{pmatrix}
0\\
1\\
\zero
\end{pmatrix}\,.
\end{align}
\end{small}
By solving
\begin{align}
\label{systemExpand2}
(\II - \WW^{\top})(\zz_{S}^{\top} \,\, \zz_{F}^{\top} )^{\top}= (-\ppi_{s'_0}^{-1} \,\, \ppi_{s'_1}^{-1}\,\, \zero^\top)^{\top}
\end{align}
 we obtain a $\zz_{F}$ that satisfies the latter $n-2$ equations in (\ref{systemExpand}). We note that (\ref{systemExpand2}) has solutions because $(-\ppi_{s'_0}^{-1} \,\, \ppi_{s'_1}^{-1}\,\, \zero^\top)^{\top}\in \ker(\II - \WW^{\top})$. Since the rank of $(\II -\WW^{\top})$ is $n-1$ and $(\II - \WW^{\top})\one = \zero$, for any $\zz$ satisfying the system of equations (\ref{systemExpand2}), $\yy = \zz+ \gamma \one$ also satisfies (\ref{systemExpand2}), where $\gamma$ can be any real number. We observe that
\begin{align}
\label{systemExpandSol1}
\zz = (\II - \WW^{\top})^{\dag} \PPi^{-1} \bb_{s'_1,s'_0}
\end{align}
is a solution of (\ref{systemExpand2}). This can be verified by plugging it into (\ref{systemExpand2}):
\begin{align*}
&(\II - \WW^{\top})(\II - \WW^{\top})^{\dag} \PPi^{-1} \bb_{s'_1,s'_0}\\
=&\PPi^{-1}\kh{ \PPi(\II - \WW^{\top})}\kh{(\II - \WW^{\top})^{\dag} \PPi^{-1}} \bb_{s'_1,s'_0}\\
=& \PPi^{-1}(\II - \frac{1}{n}\one\one^{\top}) \bb_{s'_1,s'_0}= \PPi ^{-1} \bb_{s'_1,s'_0}
\end{align*}
The second equality follows from Lemma~\ref{MatrixEqn4}. Then, we further set $\zz' = \zz - \zz_{s'_0}\one$ to make $\zz'_{s'_0} = 0$. Now we have found $\zz'$ which satisfies (\ref{systemExpand}) except for the second equation. We note that by multiplying a factor $\beta$ to $\zz'$, the other $n-1$ equations are still satisfied. So we let $\yy' = (\zz_u - \zz_v)^{-1} \zz'$. Then $\hat{\xx} = \yy'$ is the solution of (\ref{systemExpand}).
\end{IEEEproof}
}

\subsection{Proof of Theorem~\ref{partialBalance.thm}}
\label{appendix4.sec}
\begin{IEEEproof}
According to the Sherman-Morrison formula,
\begin{align}
\label{invUpdate.eqn}
&\hat{\xx}_v  = \ee_v^{\top} \Bigg(\kh{\LL + \EE^{s_0}\kappa_0}^{-1}  \nonumber\\
&\quad -\frac{\kappa_1\kh{\LL + \EE^{s_0}\kappa_0}^{-1}\EE^{s_1}\kh{\LL + \EE^{s_0}\kappa_0}^{-1}}{ 1 + \kappa_1 \ee_{s_1} ^{\top}\kh{\LL + \EE^{s_0}\kappa_0}^{-1} \ee_{s_1}}\Bigg) \ee_{s_1} \kappa_1\,.
\end{align}
Let us then consider $\kh{\LL + \EE^{s_0}\kappa_0}^{-1}$. Since $\LL$ is a singular matrix, the Sherman-Morrison formula cannot be applied in this case. Instead we apply the rank-$1$ update given in~\cite{Me73}. 
\rvs{By further applying some matrix identities discussed in Appendix~\ref{appendix1.sec}, we obtain}
\rvs{
\begin{align}
\label{psUpdate.eqn}
\kh{\LL + \EE^{s_0}\kappa_0}^{-1}= &\LL^\dag\! - \frac{1}{\qq_{s_0}}\cdot(\LL^\dag \ee_{s_0})\qq^\top\! - \one (\ee_{s_0}^\top \LL^\dag) \nonumber\\
&+ (1/\kappa_0+\ee_{s_0}^\top \LL^\dag \ee_{s_0})\cdot\frac{1}{\qq_{s_0}}\cdot\one\qq^\top \,,
\end{align}
where $\qq = \DD^{-1}\ppi$. Plugging (\ref{psUpdate.eqn}) into (\ref{invUpdate.eqn}), we arrive at
\begin{align}
\hat{\xx}_v\! =\! \frac{\frac{1}{\kappa_0\qq_{s_0}} + \bb_{v,s_0}^\top\LL^\top\DD\PPi^{-1}\bb_{s_1,s_0}}{\frac{1}{\kappa_0\qq_{s_0}} + \frac{1}{\kappa_1\qq_{s_1}} + \bb_{s_1,s_0}^\top\LL^\top\DD\PPi^{-1}\bb_{s_1,s_0}}\,.
\end{align}
We further note that for any $\pp\perp\one$, $\rr\perp\one$, $\pp^{\top}\LL^{\dag}\DD\PPi^{-1}\rr = \pp^{\top}\Lcal^{\dag}\rr$ (see Appendix~\ref{appendix1.sec} for details). Then we obtain
\begin{align}
\mu(S_1)\! = \! \frac{\frac{1}{\kappa_0\qq_{s_0}} + \ee_{s_0}^\top\Lcal^\dag\bb_{s_0,s_1}}{\frac{1}{\kappa_0\qq_{s_0}} + \frac{1}{\kappa_1\qq_{s_1}} + \bb_{s_1,s_0}^\top\Lcal^\dag\bb_{s_1,s_0}}\,,\nonumber
\end{align}
which directly leads to the desired result.
}
\end{IEEEproof}

\subsection{Proof of Theorem~\ref{hardnessP1.thm}}
\label{appendix3.sec}

\begin{problem}[\underline{V}ertex \underline{C}over on \underline{$3$} Regular Graphs]
Given an undirected connected $3$-regular graph $G=(V,E)$ and an integer $k$, decide whether there is a vertex set $S_1\subseteq V$, such that $|S_1|\leq k$ and $|S_1|$ is a vertex cover of graph $G$.
\end{problem}

We give a decision version of Problem \ref{stubborn.prob} as follows.
\begin{problem}[\rvs{\underline{A}bsolute}  \underline{L}eader \underline{S}election \underline{D}ecision Problem]
\label{stubborn2.prob}
In an \rvs{absolute leader system}, given a strongly connected directed graph $\Gcal=(\Vcal, \Ecal, w)$, an opinion $0$ leader set $S_0\neq \emptyset$, two real numbers $\alpha,\beta \in [0,1]$, a candidate set $Q\subseteq \Vcal\backslash S_0$, $|Q|=q$, and an integer $1\leq k\leq q$, decide whether there is a leader set $S_1\subseteq Q$ with opinion $1$ with at most $k$ nodes, such that the average opinion of all nodes (including leaders and followers) $\mu(S_1) = \frac{1}{n}\sum_v \hat{\xx}_v$  satisfies $f(S_1,\alpha)=|\mu(S_1)-\alpha| \leq \beta$.
\end{problem}

\begin{lemma}
\label{decisionHard.lemma}
Given an instance of problem \ref{stubborn2.prob}, it is \textbf{NP}-hard to decide if there is a set $S_1$, $|S_1|\leq k$, such that $|\mu(S_1)-\alpha|\leq \beta $
\end{lemma}
\begin{IEEEproof}
\rvs{In this proof, we consider undirected graphs.} Let $\Fcal = (\Vcal, \Ecal, w)$ be a graph consisting of a star graph $\Scal_n$ plus a $3$-regular subgraph $\Fcal[V]=(V,E,\omega)$ supported on $n-1$ leaves of $\Scal_n$. Edges in $\Scal_n$ are weighted $3$ and edges in $\Fcal[V]$ are weighted $1$. Then, we can construct an instance of Problem \ref{stubborn.prob} by letting $S_0=\{s_0\}$ be the central node of $\Scal_n$, and the candidate set $Q$ be the node set $V=\Vcal\backslash\{s_0\}$. and $k$ be any integer that satisfies $1\leq k \leq q$. 

\noindent \textbf{Completeness:} If $|S_1| = k$ and $S_1$ is a vertex cover of the $3$-regular graph $G= \Fcal[V]$, then we consider the steady-state of the followers $\hat{\xx}_F$ given by (\ref{steady.eqn}). In this case, $\hat{\xx}_{F} = \rm{diag}\kh{[6, \ldots, 6]}^{-1}[3, \ldots, 3]^{\top}=[\frac{1}{2}, \ldots, \frac{1}{2}]^{\top}$. There are $n-1-k$ follower nodes; thus, we have $\mu(S_1)=\frac{1}{n}\kh{\frac{1}{2}\cdot(n-1-k)+k}= \frac{n-1+k}{2n}$.

\noindent \textbf{Soundness:} If $S_1$ is not a vertex cover of graph $G$, then the follower node set is not an independent set. So, the matrix $\LL_{F,F}$ is a block diagonal matrix with each block associated with a connected component of graph $G[V\backslash S]$. Let $T \subseteq V\backslash S$, $|V|\geq 1$ be the node set of a connected component. 
%
%
Following the analysis given in the proof of~\cite[Theorem 4.1]{GTT13}, we obtain 
$\hat{\xx}_{u} < \frac{1}{2}$
for any $u\in T$. Then
\begin{align*}
\mu(S_1) <  \frac{n-1+k}{2n}\,.
\end{align*}

Next, we give a polynomial reduction form VC3 to ALSD: $p: \{G=(V,E),k\} \to \{\Gcal=(\Vcal, \Ecal, w), Q, k, \alpha, \beta \}$. For any given 3-regular graph $G$ with $n-1$ nodes, we construct a weighted graph $\Gcal= G + \Scal_n$, with all edges in the original graph weighted $1$ and all edges in the star $\Scal_n$ weighted $3$. Let $Q = V$, $k$ be the same integer, $\alpha$ be any constant $t$ greater or equal to $c = \frac{n-1+k}{2n}$, and $\beta  = t - c$. Then $p(G=(V,E),k) = (G+\Scal_n, V, k, t, t-c)$ is a reduction from VC3 to ALSD.
%
\end{IEEEproof}

Lemma \ref{decisionHard.lemma} immediately implies Theorem~\ref{hardnessP1.thm}.

\subsection{Proof of Theorem~\ref{algBound.thm}}
\label{appendix5.sec}

\begin{IEEEproof}
We let $\check{b}$ be the best bound found by the algorithm with the smallest $|\mu(S_{\check{b}})-\alpha|$. And, ${\rm{Greedy}}$ with the best upper bound $b^*$ returns the result $\mu(S_{b^*})$. $b^*$ is given by (\ref{bestBound.def}). 
%

If $\mu(P\cup\{s\})\leq \alpha$ is always satisfied during the execution, then $\mu(P\cup\{s\})\leq \hat{b}$ is also always satisfied. Then the returned $S_{\check{b}}$ is the same as what we get from a greedy algorithm which adds the element with largest marginal gain to the current set in each iteration until the cardinality constraint is violated. 
We further define
\begin{align*}
\widetilde{S} \in  \argmax_{T\subseteq Q,\, |T|\leq k} \mu(T)\,,
\end{align*}
therefore by the result in~\cite{NWF78} we obtain
\begin{align*}
\mu(\widetilde{S}) \geq \mu(S_{\check{b}})\geq \kh{1-\frac{1}{e}}\mu(\widetilde{S}). 
\end{align*}
If $\alpha \geq \mu(\widetilde{S})$, then $\mu(\widetilde{S}) = \mu(S^*)$, we attain the guarantee $\mu(S_{\check{b}}) \Approx{\gamma} \mu(\widetilde{S})$, where $e^{-\gamma} = (1-1/e)$. 
If $\mu(S_{\check{b}}) \leq \alpha \leq \mu(\widetilde{S})$, then $\mu(S^*)\in[\mu(S_{\check{b}}), \mu(\widetilde{S})]$, which implies ${\mu(S_{\check{b}}) \Approx{\gamma} \mu(S^*)}$, where $e^{-\gamma} = (1-1/e)$.


If $\mu(P\cup\{s\})\leq \alpha$ is first violated when we add the $(t+1)$th node, we define $P_t$ as the set of chosen nodes of size $t$ in ${\rm{Greedy}}$, therefore $|P_t|=t$. We further define $\rho(s_{t+1}) = \mu(P_t\cup\{s_{t+1}\}) - \mu(P_t)$. From the submodularity of $\mu(S)$ we know $\rho(s_{t+1})\leq \frac{1}{t+1} \mu(P_{t}\cup\{s_{t+1}\})$ holds for the greedy algorithm. Then $\mu(P_t) = \mu(P_{t}\cup\{s_{t+1}\}) - \rho(s_{t+1}) \geq \frac{t}{t+1}\mu(P_{t}\cup\{s_{t+1}\}) \geq  \frac{k_{\mu,\alpha} -1}{k_{\mu,\alpha}}\mu(P_{t}\cup\{s_{t+1}\})$. By letting $\bar{b} = \mu(P_{t}\cup\{s_{t+1}\})$ (then by definition $\bar{b}=\mu(S_{\bar{b}})=\mu(P_{t}\cup\{s_{t+1}\})$), we obtain $\mu(S_\alpha) \geq (1-\frac{1}{k_{\mu,\alpha}})\mu(S_{\bar{b}})$.
We further attain $\mu(S^*),\mu(S_{b^*}) \in [\mu(S_{\alpha}), \mu(S_{\bar{b}})]$, and $b^*\in[\alpha, \bar{b}]$. Since $\check{b},b^*\in[b_{\min},b_{\max}]$ and $b_{\min}\Approx{\delta}b_{\max}$, $\check{b}\Approx{\delta}b^*$ we obtain $\check{b}\in[\alpha, e^{\delta}\bar{b}]$ and therefore $\mu(S_{\check{b}}) \in [\mu(S_\alpha), e^{\delta}\mu(S_{\bar{b}})]$, so $\mu(S_{\check{b}}) \Approx{\gamma}  \mu(S^*)$, where $e^{-\gamma} = (1-1/k_{\mu,\alpha}) e^{-\delta}$.
%
\end{IEEEproof}

\subsection{Proof of Theorem~\ref{complexity.thm}}
\label{appendix0.sec}
\begin{IEEEproof}
We take the algorithm for \rvs{the \CSL} Selection problem as an example. 
In each execution of Line 3 of Algorithm~\ref{alg:Greedy}
, we need to calculate the sum of steady states of followers given by
 $-\one^{\top} \kh{\LL_{F, F}}^{-1} \LL_{F, S}\xx_{S}$,
 for all $P\cup \{u\}$, $u\in Q$. $P$ and $Q$ are the current leader set of opinion $1$ and the current candidate set. 
Calculating $(\LL_{F,F})^{-1}$ when $S_1=\emptyset$ takes $O(n^3)$ running time. $\LL_{FF}$ can be updated at iteration $t+1$ by deleting the row and column associated with candidate node $u$. From block matrix inversion, we obtain that its inverse can be updated by
\begin{align*}
&\kh{\LL_{(F(t)\backslash\{u\}), (F(t)\backslash\{u\})}}^{-1} =\Bigg( \kh{\LL_{F(t),F(t)}}^{-1} \nonumber\\
&\,\,\,\,- \frac{\kh{\LL_{F(t), F(t)}}^{-1} \ee_{u}\ee_{u}^\top \kh{\LL_{F(t),F(t)}}^{-1}}{\ee_{u}^\top \kh{\LL_{F(t),F(t)}}^{-1} \ee_{u}} \Bigg)_{(F(t)\backslash\{u\}), (F(t)\backslash\{u\})}\,.
\end{align*}

To calculate $\mu(P_t)$, we do not need to find $\kh{\LL_{(F(t)\backslash\{u\}), (F(t)\backslash\{u\})}}^{-1}$ explicitly.  It suffices to compare
$
-\one^{\top} \kh{\LL_{(F(t)\backslash\{u\}), (F(t)\backslash\{u\})}}^{-1} \LL_{(F(t)\backslash\{u\}), (S(t)\cup \{u\})}\xx_{S\cup \{u\}}\,,
$
for all $u$ in the current candidate set.
We note that $\ee_u$ ($\ee_u^\top$) takes a column (row) of $\kh{\LL_{F(t),F(t)}}^{-1}$, and $\LL_{(F(t)\backslash\{u\}), (S(t)\cup \{u\})}\xx_{S\cup \{u\}}$ is a column vector. By the associative law, we compute the vector inner product first and  find the updated $\mu(S_1)$ for at most $n$ candidates in $O(n^2)$ total running time. The operations of taking the submatrices do not change the complexity because for any candidate $u$, these operations only take \rvs{$O(|\Ncal^\uparrow_u| + |\Ncal^\downarrow_u|)$} running time. 
So, in each execution of Line 3 of Algorithm~\ref{alg:Greedy}, these operations can be done in $O(m)$ total running time, where $m$ is the number of edges in the graph.  After we find the best choice $s_{t+1}$ in step $t+1$, we update $(\LL_{F,F})^{-1}$ explicitly, which takes additional $O(n^2)$ time. Therefore, execution of Line 3 of Algorithm~\ref{alg:Greedy} takes $O(n^2)$ time. By using this simple acceleration, the complexity of Algorithm~\ref{alg:Greedy} is improved to $O(n^3+kn^2)= O(n^3)$. Algorithm~\ref{alg:boundSearch} calls ${\rm{Greedy}}$ $O(\log\frac{1}{\delta})$ times until $b_{\max} \Approx{\delta} b_{\min}$. Since $b_{\max}-b_{\min}$ decreases geometrically in Algorithm~\ref{alg:boundSearch}, the total running time of $\rm{BoundSearch}$ is $O(n^3\log{\frac{1}{\delta}})$. 

For the \rvs{\PSL} Selection Problem, the the rank-$1$ update is obtained using the Sherman-Morrison formula. And, the running time of the ${\rm{Greedy}}$ routine is also $O(n^3)$ by a similar implementation. We omit the details of the analysis.
\end{IEEEproof}

\subsection{Monotonicity and Submodularity of $\mu(S_1)$}
\label{appendix7.sec}
We present simple proofs for the submodularity based on the escape probability interpretation of $\hat{\xx}$.
\subsubsection{Steady-State Opinion Interpreted as Escape Probability}


The entries of $\hat{\xx}_F$ can be interpreted as the escape probability of a random walker~\cite{DS00} in a Markov chain with absorbing states define on graph $\Gcal$.
%
%
 Consider an absorbing Markov chain $\PP$ with $S_0\cup S_1$ the set of absorbing states and $F$ the set of non-absorbing states. Then the transition matrix has the form
\begin{align}
\label{absorbChain.def}
\PP^\top = \begin{pmatrix}
\II & \zero\\
\RR & \QQ
\end{pmatrix}\,.
\end{align}
where $\RR = (\DD_{F,F})^{-1} \AA_{F,S}$ and $\QQ = (\DD_{F,F})^{-1} \AA_{F,F}$. 

Define a harmonic function $\yy$ with boundary $\yy_B = \hat{\xx}_S$. The interior $\yy_D$ is determined by~(see \cite{DS00}, for a similar formulation for undirected graphs) 
\begin{align}
\yy_D &= \kh{\II - \QQ}^{-1} \RR \yy_B\,. 
\end{align}
Then we obtain
\begin{align*}
\hat{\xx}_F = \yy_D = -\kh{\LL_{F,F}}^{-1}\LL_{F,S}\xx_S\,.
\end{align*}
Combining with the boundary condition $\yy_B = \hat{\xx}_S$, we obtain  $\yy = \hat{\xx}$. $\yy$ defines the concept of escape probability explained below.

Let $S_0$ and $S_1$ be two sets of absorbing states in a Markov chain (\ref{absorbChain.def}). We let $\tau_v^{\Gcal}(S_1, -S_0)$ represent the event that in a Markov chain defined by graph $\Gcal$, a random walker starts from node $v$, hits any state $u\in S_1$ before it reaches any state $u\in S_0$. Then $\hat{\xx}_v$ is the probability that $\tau_v^\Gcal(S_1, -S_0)$ happens. We denote the escape probability as $p^{\Gcal}_v(S_1,-S_0)\defeq \pr{\tau_v^\Gcal(S_1,-S_0)}$. 
This escape probability is given by the harmonic function $\yy$ defined above (for example, see~\cite{DS00}). We have shown that $\yy = \hat{\xx}$, so $\hat{\xx}_v = p^{\Gcal}_v(S_1,-S_0)$. Similarly, we define $\tau_v^\Gcal(S_0, -S_1)$ as the event that in the Markov chain defined by graph $\Gcal$, a random walker starts from node $v$, hits any state $u\in S_0$ before it reaches any state $u\in S_1$, and we also denote by $p^{\Gcal}_v(S_0,-S_1)\defeq \pr{\tau_v^\Gcal(S_0,-S_1)}$ the probability that event $\tau_v^\Gcal(S_0, -S_1)$ happens.
Since a random walker is either absorbed by $u\in S_0$ or $u\in S_1$, $p^{\Gcal}_v(S_1,-S_0) + p^{\Gcal}_v(S_0,-S_1)=1$.



\subsubsection{\rvs{Internal Leader System}}
In the considered leader-follower system with \rvs{\csl}s, given fixed $S_0$, $\mu(S_1)$ is defined as
\begin{align}
\mu(S_1)\defeq \frac{1}{n}\sum_{v\in \Vcal} p^{\Gcal}_v(S_1, -S_0)\,.
\end{align}
To prove that $\mu(S_1)$ is monotone and submodular, it suffices to show that $p^\Gcal_v(S_1, -S_0)$ is monotone and submodular for all $v\in \Vcal$.
\begin{lemma}
\label{completeMono.lemma}
For any $S_1\subseteq T_1 \subseteq \Vcal$, $S_0\subseteq \Vcal$, and $T_1 \cap S_0 = \emptyset$, for any $v\in \Vcal$
$$p^\Gcal_v( T_1, -S_0)\geq p^\Gcal_v(S_1 , -S_0)$$
\end{lemma}
\begin{IEEEproof}
We first consider $S_1^{(0)} = S_1$ and $S_1^{(1)} = S_1\cup \{u\}$, where $u\in (T_1\backslash S_1)$. For a random walker in graph $\Gcal$ starting from node $v$, we observe that
\begin{align}
& p^{\Gcal}_{v}(S_1^{(1)}, -S_0) = p^{\Gcal}_{v}((S_1\cup \{u\}), -S_0) \nonumber\\
& = p^{\Gcal}_{v}(S_1, -(S_0\cup \{u\})) + p^{\Gcal}_{v}(\{u\}, -(S_0\cup S_1))
\end{align}
and
\begin{align}
& p^{\Gcal}_{v}(S_1^{(0)}, -S_0)  = p^{\Gcal}_{v}((S_1, -S_0) \nonumber\\
&= p^{\Gcal}_{v}(S_1, -(S_0\cup \{u\})) \nonumber\\
&\quad + p^{\Gcal}_{v}(\{u\}, -(S_0\cup S_1))\cdot p^{\Gcal}_{u}(S_1, -S_0)\,,
\end{align}
by the Markov property. Therefore
\begin{align}
&p^{\Gcal}_{v}(S_1^{(1)}, -S_0) - p^{\Gcal}_{v}(S_1^{(0)}, -S_0) \nonumber\\
&= p^{\Gcal}_{v}(\{u\}, -(S_0\cup S_1))\cdot \kh{ 1-p^{\Gcal}_{u}(S_1, -S_0)} \nonumber\\
& = p^{\Gcal}_{v}(\{u\}, -(S_0\cup S_1))\cdot p^{\Gcal}_{u}(S_0, -S_1) \geq 0\,.
\end{align}
Similarly, by defining a sequence of $S_1^{(i)}, i = 1, \ldots, t$ such that $t = |T_1\backslash S_1|$ and $S_1^{(t)} = T_1$, we attain
\begin{align}
p^{\Gcal}_{v}(S_1^{(i)}, -S_0) \geq  p^{\Gcal}_{v}(S_1^{(i-1)}, -S_0)
\end{align}
holds for all $i\in [t]$. And this leads to the result in lemma~\ref{completeMono.lemma}.
\end{IEEEproof}

Since $p^{\Gcal}_v(S_1,-S_0) = 1-p^{\Gcal}_v(S_0,-S_1)$, we attain the following corollary
\begin{corollary}
\label{avoidT.corollary}
For any $S_0\subseteq \Vcal$, $S_1\subseteq T_1 \subseteq \Vcal$, and $T_1 \cap S_0= \emptyset$
$$p^{\Gcal}_v(S_0, -T_1) \leq p^{\Gcal}_v( S_0, -S_1)\,.$$
\end{corollary}

\begin{lemma}
For any $S_1\subseteq T_1 \subseteq \Vcal$, $S_0 \subseteq \Vcal$, $T_1\cap S_0 =\emptyset$, and $u\in \Vcal \backslash (T_1\cup S_0)$, 
\begin{align}
& p^{\Gcal}_v(T_1\cup \{u\}, -S_0)-p^{\Gcal}_v(T_1, -S_0) \nonumber\\
& \leq p^{\Gcal}_v(S_1\cup \{u\}, -S_0)-p^{\Gcal}_v(S_1, -S_0)
\end{align}
\end{lemma}
\begin{IEEEproof}
For any $v\in \Vcal$,
\begin{align}
\label{SpMarginal.eqn}
& p^{\Gcal}_v(T_1\cup \{u\}, -S_0)-p^{\Gcal}_v(T_1, -S_0) \nonumber\\
&= p^{\Gcal}_v(\{u\}, -(T_1\cup S_0)) \cdot p^{\Gcal}_u( S_0, -T_1) 
\end{align}
and
\begin{align}
\label{SMarginal.eqn}
& p^{\Gcal}_v(S_1\cup \{u\}, -S_0)-p^{\Gcal}_v(S_1, -S_0) \nonumber\\
&= p^{\Gcal}_v(\{u\}, -(S_1\cup S_0)) \cdot p^{\Gcal}_u(S_0, -S_1)
\end{align}
Using corollary \ref{avoidT.corollary} we get the inequality in the lemma by comparing (\ref{SpMarginal.eqn}) and (\ref{SMarginal.eqn}).

\end{IEEEproof}


\subsubsection{\rvs{External Leader System}}
In the considered leader-follower system with \rvs{\psl}s, given fixed $S_0$, $\mu(S_1)$ is defined as
\begin{align}
\mu(S_1)\defeq \frac{1}{n}\sum_{v\in \Vcal} p^{\Gcal'}_v(\{s'_1\}, -\{s'_0\})\,,
\end{align}
in which $p^{\Gcal'}_v(\{s'_1\}, -\{s'_0\})$ represents the probability that a random walker in augmented graph $\Gcal'$ starting from $v$ reaches $s'_1$ before it reaches $s_0$. 
To prove that $\mu(S_1)$ is monotone and submodular, it suffices to show that $p^{\Gcal'}_v(\{s_1\}, -\{s_0\})$ is monotone and submodular for all $v\in \Vcal$.
\begin{lemma}
\label{partialMono.lemma}
For any $S_1\subseteq T_1 \subseteq \Vcal$, $S_0\subseteq \Vcal$, and $T_1 \cap S_0 = \emptyset$,
we consider the augmented graph $\Gcal'$ defined by $\Gcal$, $S_0$, and $S_1$; and the augmented graph $\Hcal'$ defined by $\Gcal$, $S_0$ and $T_1$. Then $\Hcal'$ has the same node set as $\Gcal'$, the edge set of $\Hcal'$ consists of all edges in the edge set of $\Gcal'$, and all $(u,s_1'), u\in(T_1\backslash S_1)$.
 For any $v\in \Vcal$
$$p^{\Hcal'}_v( \{s'_1\}, -\{s'_0\})\geq p^{\Gcal'}_v(\{s'_1\} , -\{s'_0\})\,.$$
\end{lemma}
\begin{IEEEproof}
Let $\Gcal+(u,v)$ be the graph attained by adding an edge $(u,v)$ to the graph $\Gcal$. We start by considering $\Gcal^{(0)}= \Gcal'$ and $\Gcal^{(1)} = \Gcal' + (u, s'_1)$, $u \in (T_1\backslash S_1)$. Let $\xi^{\Gcal'}_v(u, s'_1)$ be the event that a random walker in $\Gcal'$ starting from node $v$ passes through edge $(u,s'_1)$ before it reaches any absorbing state, and $\overline{\xi}^{\Gcal'}_v(u, s'_1)$ be the event that a random walker does not pass through $(u,s'_1)$ before reaching an absorbing state.
\begin{align*}
&p^{\Gcal^{(1)}}_v(\{s'_1\},-\{s'_0\})  \nonumber\\
=&  \pr{\tau_v^{\Gcal^{(1)}}(\{s'_1\}, -\{s'_0\}) \Big| \xi^{\Gcal^{(1)}}_v(u, s'_1) }\pr{\xi^{\Gcal^{(1)}}_v(u, s'_1)} \\
&+ \pr{\tau_v^{\Gcal^{(1)}}(\{s'_1\}, -\{s'_0\}) \Big|\overline{\xi}^{\Gcal^{(1)}}_v(u, s'_1) }\pr{\overline{\xi}^{\Gcal^{(1)}}_v(u, s'_1)}\nonumber\\
=& \pr{\xi^{\Gcal^{(1)}}_v(u, s'_1)} + \pr{\tau_v^{\Gcal^{(1)}}(\{s'_1\}, -\{s'_0\}) \Big| \overline{\xi}^{\Gcal^{(1)}}_v(u, s'_1) }\nonumber\\
&\qquad\qquad\qquad\qquad\qquad\qquad \cdot  \kh{ 1 - \pr{\xi^{\Gcal^{(1)}}_v(u, s'_1)}}\,.
\end{align*}
We note that 
\begin{align*}
&\pr{\tau_v^{\Gcal^{(1)}}(\{s'_1\}, -\{s'_0\}) \Big| \overline{\xi}^{\Gcal^{(1)}}_v(u, s'_1) } \nonumber\\
&= \pr{\tau_v^{\Gcal^{(0)}}(\{s'_1\}, -\{s'_0\})}\,,
\end{align*}
therefore
\begin{align}
& p^{\Gcal^{(1)}}_v(\{s'_1\},-\{s'_0\}) - p^{\Gcal^{(0)}}_v(\{s'_1\},-\{s'_0\}) \nonumber\\
&= \pr{\xi^{\Gcal^{(1)}}_v(u, s'_1)}\cdot \kh{ 1  - p^{\Gcal^{(0)}}_v(\{s'_1\},-\{s'_0\}) }\nonumber\\
& = \pr{\xi^{\Gcal^{(1)}}_v(u, s'_1)} \cdot p^{\Gcal^{(0)}}_v(\{s'_0\},-\{s'_1\}) \geq 0\,.
\end{align}
Similarly, by defining a sequence of $\Gcal^{(i)}$, $i=1,\ldots, t$ such that $t = |T_1\backslash S_1|$, we attain $\Gcal^{(t)} = \Hcal'$ and $$p^{\Gcal^{(i)}}_v(\{s'_1\},-\{s'_0\}) \geq  p^{\Gcal^{(i-1)}}_v(\{s_1\},-\{s_0\})$$ holds for all $i\in[t]$. This leads to the result in lemma~\ref{partialMono.lemma}.
\end{IEEEproof}
Since $P^{\Gcal'}_v(\{s'_1\},-\{s'_0\}) = 1-P^{\Gcal'}_v(\{s'_0\},-\{s'_1\})$, we obtain the following corollary
\begin{corollary}
\label{avoidT2.corollary}
For any $S_0\subseteq \Vcal$, $S_1\subseteq T_1 \subseteq \Vcal$, and $T_1 \cap S_0= \emptyset$, $\Gcal'$ and $\Hcal'$ have the same definitions as they are defined in Lemma~\ref{partialMono.lemma}. then
$$p^{\Hcal'}_v( \{s'_0\}, -\{s'_1\})\leq p^{\Gcal'}_v(\{s'_0\} , -\{s'_1\})\,.$$
\end{corollary}

\begin{lemma}
\label{partialSubm.lemma}
For any $S_1\subseteq T_1 \subseteq \Vcal$, $S_0\subseteq \Vcal$, $T_1 \cap S_0 = \emptyset$, and $u\notin (S_0\cup T_1)$, 
we consider the augmented graph $\Gcal'$ defined by $\Gcal$, $S_0$, and $S_1$; and the augmented graph $\Hcal'$ defined by $\Gcal$, $S_0$ and $T_1$. Then $\Hcal'$ has the same node set as $\Gcal'$, the edge set of $\Hcal'$ consists of all edges in the edge set of $\Gcal'$, and all $(l,s_1'), l\in(T_1\backslash S_1)$.
 For any $v\in \Vcal$ and $u\notin (S_0\cup T_1)$,
\begin{align}
&p^{\Hcal'+ (u,s'_1)}_v( \{s'_1\}, -\{s'_0\}) - p^{\Hcal'}_v( \{s'_1\}, -\{s'_0\}) \nonumber\\
&\leq p^{\Gcal'+(u,s'_1)}_v(\{s'_1\} , -\{s'_0\})- p^{\Gcal'}_v(\{s'_1\} , -\{s'_0\})\,.
\end{align}
\end{lemma}

\begin{IEEEproof}
Following similar analysis as the proof of Lemma~\ref{partialMono.lemma}, we obtain
\begin{align}
\label{SpMarginal2.eqn}
&p^{\Hcal'+ (u,s'_1)}_v( \{s'_1\}, -\{s'_0\}) - p^{\Hcal'}_v( \{s'_1\}, -\{s'_0\}) \nonumber\\
&= \pr{\xi^{\Hcal'+(u,s'_1)}_v(u, s'_1)} \cdot p^{\Hcal'}_v(\{s'_0\},-\{s'_1\})
\end{align}
\begin{align}
\label{SMarginal2.eqn}
&p^{\Gcal'+ (u,s'_1)}_v( \{s'_1\}, -\{s'_0\}) - p^{\Gcal'}_v( \{s'_1\}, -\{s'_0\}) \nonumber\\
&= \pr{\xi^{\Gcal'+(u,s'_1)}_v(u, s'_1)} \cdot p^{\Gcal'}_v(\{s'_0\},-\{s'_1\})
\end{align}
Then we extend the definition of $\xi^{\Gcal'}_v(u, s'_1)$ and denote $\xi^{\Gcal'}_v(U, s'_1)$ as the event that a random walker in $\Gcal'$ starting from $v$ passes through any edge $(u,s'_1)$, $u\in U$ before it reaches any absorbing state. Similarly we define $\overline{\xi}^{\Gcal'}_v(U, s'_1)$ as the event that the random walker reaches an absorbing state without passing through any $(u,s'_1)$, $u\in U$.
\begin{align}
&\pr{\xi^{\Hcal'+(u,s'_1)}_v(u, s'_1)} \nonumber\\
&= \pr{\xi^{\Hcal'+(u,s'_1)}_v(u, s'_1) \Big| \overline{\xi}^{\Hcal'+(u,s'_1)}_v((T_1\backslash S_1), s'_1)}\nonumber\\
&\qquad\qquad\qquad\qquad \cdot \pr{\overline{\xi}^{\Hcal'+(u,s'_1)}_v((T_1\backslash S_1), s'_1)}\nonumber\\
&+ \pr{\xi^{\Hcal'+(u,s'_1)}_v(u, s'_1) \Big| \xi^{\Hcal'+(u,s'_1)}_v((T_1\backslash S_1), s'_1)}\nonumber\\
&\qquad\qquad\qquad\qquad \cdot \pr{\xi^{\Hcal'+(u,s'_1)}_v((T_1\backslash S_1), s'_1)}
\end{align}
In addition, $$\pr{\xi^{\Hcal'+(u,s'_1)}_v(u, s'_1) \Big| \xi^{\Hcal'+(u,s'_1)}_v((T_1\backslash S_1), s'_1)} = 0$$ and 
\begin{align}
&\pr{\xi^{\Hcal'+(u,s'_1)}_v(u, s'_1) \Big| \overline{\xi}^{\Hcal'+(u,s'_1)}_v((T_1\backslash S_1), s'_1)} \nonumber\\
&=  \pr{\xi^{\Gcal'+(u,s'_1)}_v(u, s'_1)}\,,
\end{align}
then we obtain
\begin{align}
\label{probDec.ineq}
\pr{\xi^{\Hcal'+(u,s'_1)}_v(u, s'_1)} \leq \pr{\xi^{\Gcal'+(u,s'_1)}_v(u, s'_1)}
\end{align}
Applying Corollary \ref{avoidT2.corollary} and (\ref{probDec.ineq}) to (\ref{SpMarginal2.eqn}) and (\ref{SMarginal2.eqn}) leads to the result stated in Lemma \ref{partialSubm.lemma}.
\end{IEEEproof}

\end{document}